\documentclass[12pt,english, journal,draftclsnofoot,onecolumn]{IEEEtran}
\usepackage[T1]{fontenc}
\usepackage[latin9]{inputenc}
\usepackage{geometry}
\usepackage{color}
\usepackage{babel}
\usepackage{textcomp}
\usepackage{mathrsfs}
\usepackage{amsthm}
\usepackage{amsmath}
\usepackage{amssymb}
\usepackage{stackrel}
\usepackage{graphicx}
\usepackage{setspace}
\usepackage{esint}
\usepackage[unicode=true,pdfusetitle,
bookmarks=true,bookmarksnumbered=false,bookmarksopen=false,
breaklinks=false,pdfborder={0 0 1},backref=false,colorlinks=false]
{hyperref}
\usepackage{breakurl}

\makeatletter

\providecommand{\tabularnewline}{\\}

\usepackage{enumitem}		
\theoremstyle{plain}
\newtheorem{thm}{\protect\theoremname}

\usepackage{colortbl}
\usepackage{cite}
\usepackage{nomencl}
\usepackage{flushend}
\usepackage{stfloats}
\usepackage[margin=8pt,font=footnotesize,labelfont=md,labelsep=colon]{caption}

\@ifundefined{showcaptionsetup}{}{%
	\PassOptionsToPackage{caption=false}{subfig}}
\usepackage{subfig}
\makeatother

\providecommand{\theoremname}{Theorem}

\begin{document}

\title{Rethinking Dense Cells for Integrated Sensing and Communications:
A Stochastic Geometric View}

\author{Abdelhamid Salem,\textit{\normalsize{} Member, IEEE}{\normalsize{},} Kaitao Meng,\textit{\normalsize{} Member, IEEE},
Christos Masouros, \textit{\normalsize{}Senior Member, IEEE}, Fan
Liu\textit{\normalsize{}, Member, IEEE}{\normalsize{},} and David
Lopez-Perez, \textit{\normalsize{}Senior Member, IEEE} {\normalsize{}}\\
\thanks{Abdelhamid Salem, Kaitao Meng, and Christos Masouros are with the Department of
Electronic and Electrical Engineering, University College London,
London, UK, (emails: \{a.salem, kaitao.meng, c.masouros\}@ucl.ac.uk). Abdelhamid Salem is also affiliated with Benghazi University, Benghazi, Libya. Fan Liu is
with the Department of Electronic and Electrical Engineering, Southern
University of Science and Technology, Shenzhen 518055, China, (e-mail:
liuf6@sustech.edu.cn). David Lopez-Perez is with the Algorithm and
Software Design Department, Huawei Technologies, 92100 Boulogne-Billancourt,
France (e-mail: dr.david.lopez@ieee.org).%
}}
\maketitle
\begin{abstract}
\textcolor{blue}{The inclusion of the sensing functionality in the
coming generations of cellular networks, necessitates a rethink of
dense cell deployments. In this paper, we analyze and optimize dense
cell topologies for dual-functional radar-communication (DFRC) cellular
networks. With the aid of tools from stochastic geometry, we derive
new analytical expressions of the potential area spectral efficiencies
in ($\textrm{bit}/\textrm{sec}/m^{2}$) of radar and communication
systems. Based on the new formulations of the potential area spectral efficiencies,
the energy efficiency (bit/Joule) of DFRC systems is provided in a closed-form formula. Then, an optimization problem to obtain
the optimal base station (BS) density that maximizes the network-level
energy efficiency is formulated and investigated. In this regard,
the mathematical expression of the energy efficiency is shown to be
a uni-modal and pseudo-concave function in the density of the BSs.
Therefore, the optimal density of the BSs that maximizes the energy efficiency
can be obtained. Our analytical and numerical results demonstrate
that the inclusion of the sensing functionality clearly differentiates
the optimal BS topologies for the DFRC systems against classical communication-only
systems.}\end{abstract}


\section{Introduction}

Next-generation wireless networks will involve much more beyond communications,
to provide new functionalities including sensing, localization, and activity
detection. The integration of sensing and communication (ISAC) has
been recognized as a key 6G technology \cite{Generalized,rad2,joint,Analysis,Hybrid,Spectral,OFDM,survy,fannew,fanew2,meng2023throughput,imanpaper}.
To reduce costs and improve spectral and energy efficiencies, Dual-functional
Radar-Communication (DFRC) technique has received considerable attention
from both industry and academia \cite{fannew,fanew2}.  DFRC technique
combines both radar sensing and wireless communications via shared
use of the spectrum, hardware platform and a signal processing framework
\cite{Generalized,rad2,joint,Analysis,Hybrid,Spectral,OFDM,survy}.
 DFRC system has been widely considered in the literature. For instance,
in \cite{Generalized} the authors considered  DFRC beam-forming design
to simultaneously detect the targets as a multiple-input multiple
output (MIMO) radar and communicate to multiple users. To characterize
the performance tradeoff between MIMO radar and multiple users communication,
the authors first defined the achievable performance region of the
DFRC system, then both radar-centric and communication-centric optimizations
were formulated to achieve the boundary of the performance region.
In \cite{rad2} an exact closed-form expression for the probability
of false alarm was derived, while probability of detection was approximated
by assuming the signal-to-noise ratio of the reference channel is
much larger than that of the surveillance channel. Further work in
\cite{joint} introduced a joint transmit beamforming model for a
dual-function MIMO radar and multiple communication users. The proposed
dual-function system transmits the weighted sum of independent radar
waveforms and communication symbols, forming multiple beams towards
the radar targets and the communication users, respectively. In \cite{Analysis}, a
closed-form expression of average ambiguity function has been firstly
obtained, then the authors proposed a joint optimization method to
improve radar peak side lobe level influenced by communication signals.
In \cite{Hybrid} the hybrid transmit/receive beam-formers have been
designed by maximizing the sum-rate under transmit power constraints
and the similarity between the designed beamformer and the one that
has good beam pattern properties. In \cite{Spectral} the authors considered
the mutual information between the target reflections and the target
responses for DFRC systems as the design metric. Accordingly, the authors
obtained the optimal waveforms with the maximum mutual information.
The authors in \cite{OFDM} considered a MIMO DFRC system, which senses
several directions and serves multiple users. Based on an orthogonal
frequency division multiplexing transmission, the design of the radiated
waveforms and of the receive filters employed by the radar and the
users have been investigated. \textcolor{blue}{In \cite{Rev1ref2}
new integrated scheduling method of sensing, communication, and control
has been presented.} A comprehensive survey of the research progress
in the areas of radar-communication coexistence and DFRC systems with
particular emphasis on application scenarios and technical approaches
was presented in \cite{survy}. While the above consider DFRC systems
on a link level, the literature is sparse on the network-level analysis
and optimization of ISAC systems.

Nevertheless, there is an abundance of literature on the network-level
performance of communication only cellular networks. In the past few
years, the modeling and analysis of wireless communication networks
have been considered by the mathematical tool of stochastic geometry,
more precisely, by the theory of spatial point processes \cite{Marco}.
From the system-level point of view, it has been empirically validated
that the locations of the base stations (BSs) can be modeled as points
of a homogeneous Poisson point process (PPP) whose intensity coincides
with the average number of BSs per unit area \cite{Marco}. Motivated
by this, the PPP modeling approach has been widely used to analyze
the trade-off between the network spectral and energy efficiencies.
In \cite{Marco} the authors analyzed the energy efficiency of down-link
cellular networks using stochastic geometry tools. Mathematical framework
of the average spectral efficiency of multi\textendash tier cellular
networks, in which single antenna BSs have been distributed in the
network according to a PPP was presented in \cite{avgrate}. The authors
in \cite{muplink} provided analytical frameworks for system-level
analysis and design of up-link heterogeneous cellular networks where
the locations of multiple antennas BSs modeled as points of homogeneous
PPP. In addition, a novel framework based on stochastic geometry for
the co-existence of aerial and terrestrial users was developed in
\cite{uavding}, where the spatial distribution of the BSs has been
assumed to follow a PPP. \textcolor{blue}{Tractable framework for
symbol error probability, outage probability, ergodic rate, and throughput
for downlink cellular networks with different MIMO configurations
based on SG approach have been provided in \cite{Rev3Ref1}}. \textcolor{blue}{In
\cite{Rev3Ref2} analytical framework to study the joint impact of
the sensor and the user equipment (UE) densities, on drones detection.}
In \cite{newsg1} a new mathematical approach that relies on a PPP
model for the BSs locations was introduced to evaluate the performance
of down-link MIMO cellular networks. In \cite{newsg2} a new mathematical
framework to compute the error probability of downlink cellular networks
based on the PPP model for the spatial BSs locations was introduced. \textcolor{blue}{New
models for the coverage and rate of cellular networks have been developed
in \cite{Marco3} using stochastic geometry.} Based on these models,
mathematical expressions for the coverage probability and the mean
rate have been derived. In \cite{Pelaez} based on a stochastic geometry
approach, new analytical frameworks to calculate the coverage probability
and average rate of cellular networks were derived with the aid of
the Gil-Pelaez inversion formula. \textcolor{blue}{A general study
of the energy and spectral efficiencies of cellular networks has been
presented in \cite{book,Rev1ref1,Rev1ref3,Rev1ref4}. }

Against the state-of-the-art of research on performance evaluation
of DFRC systems, in this work with the aid of tools from stochastic
geometry approach, we focus our attention on system-level analysis
and optimization. More specifically, by taking into account the impact
of network deployments we introduce a new mathematical framework for
DFRC cellular networks. In this regard, based on potential area spectral
efficiency (PSE), which is the network information rate per unit area
(measured in bit/sec/$m^{2}$), the energy efficiency of the DFRC network
is analyzed \cite{Marco3}. After that, an energy efficiency optimization
problem is formulated to obtain the optimal BSs density.

\textcolor{blue}{It has been empirically validated that, from the
system-level perspective, the locations of the BSs can be modeled
as points of a homogeneous PPP, whose intensity coincides with the
average number of BSs per unit area. Motivated by these results, the
PPP modeling approach for the locations of the DFRC BSs can be used
to design the ISAC networks. Furthermore, system-level analysis and
optimization are beneficial approaches when the network designers are
interested in optimizing the performance of the entire ISAC networks.
They can be used to optimize the current ISAC networks, and also to
develop and plan future networks. In addition, analyzing and designing
ISAC networks from the energy efficiency perspective necessitate proper
mathematical tools, which are different from the formulas that used
for optimizing the network spectral efficiency and the energy consumption
individually. This optimization problem, should be formulated in a
sufficiently simple realistic manner, so that all the relevant system
parameters appear explicitly. }

The novel contributions of this paper are listed as follows

1- We introduce a new analytical framework based on SG for the analysis
of both sensing and communication performance for ISAC networks;

2- We derive a new closed-form analytical formulation of the PSE in
(bit/sec/$m^{2}$) for communication systems which depend on the density
of the BSs.

3- We derive a new closed-form analytical expression of the PSE for
radar system is derived as function of the density of the BSs. \textcolor{blue}{The
derived expressions are in closed form, without any integrations or
expectations. In addition, from these expressions we can notice the
impact of the system parameters on the system performance, and can
be used to optimize and design the systems by formulating optimization
problems, which is the main aim of these expressions. }

4- Based on the new formulations of the potential area spectral efficiencies,
the energy efficiency (bit/Joule) of  DFRC systems is provided in
a closed-form formula.

5- An energy efficiency optimization problem is formulated to find
optimal BSs density that maximizes the energy efficiency of  DFRC
systems.

6- Monte-Carlo simulations are also provided to confirm the accuracy
of the analysis, and to examine, and investigate the impact of several
parameters on the system performance, and reveal the impact of the
sensing functionality on the network-level design of the ISAC networks.

\textcolor{blue}{The insights revealed in this work can be summarized as follows:}

\textcolor{blue}{
1- From the energy efficiency perspective, the optimal BS density for ISAC networks is lower than that for communication-only networks, which in turn changes the BS deployment significantly. This is expected since the sensing component can effectively compensate for the potential efficiency loss in the communication systems. }

\textcolor{blue}{
2- Achieving the optimal energy efficiency, the equivalent coverage radius on average is effectively improved, e.g., for the typical 5G system that has a radius area 250 m, yields an optimum radius $162.8675$ m for a communication-only cellular network changes to $199.4711$ m for an ISAC cellular network.}

\textcolor{blue}{
3- In the considered interference-limited scenarios, it is more energy-efficient for ISAC networks to work in low transmit power regions, i.e., less than $35$dBm. Also, the energy efficiency degrades as the height of the target increases.}

This paper is organized as follows. In section \ref{sec:System-Model},
we describe the system model. Section \ref{sec:PSE-of-RadCom} derives
analytical expressions for the PSE and energy efficiency of communication
and radar systems. Optimal BSs density is considered in Section \ref{sec:Optimal-BS-Density}.
Numerical examples and simulation results are presented and discussed
in section \ref{sec:Numerical-Results}. Finally, Section \ref{sec:Conclusion}
outlines the main conclusions of this work.

\begin{table}
\begin{centering}
\textcolor{blue}{}%
\begin{tabular}{|c|c|}
\hline 
\textcolor{blue}{$(\cdot)^{H}$ and $(\cdot)^{\dagger}$ } & \textcolor{blue}{Conjugate transposition, and transposition }\tabularnewline
\hline 
\textcolor{blue}{$\mathcal{E}\left[\cdot\right]$ and $\textrm{diag}\left(\cdot\right)$} & \textcolor{blue}{Average operation and diagonal of a matrix}\tabularnewline
\hline 
\textbf{\textcolor{blue}{$\mathbf{H}$}}\textcolor{blue}{{} and $\mathbf{h}$ } & \textcolor{blue}{Channel matrix and vector }\tabularnewline
\hline 
\textcolor{blue}{$\left[\mathbf{h}\right]_{k}$ and $h$} & \textcolor{blue}{Element $k$ in vector $\mathbf{h}$and a scalar }\tabularnewline
\hline 
\textcolor{blue}{$\left|\cdot\right|$and $\left\Vert \cdot\right\Vert ^{2}$ } & \textcolor{blue}{Absolute value and Second norm }\tabularnewline
\hline 
\textcolor{blue}{$\mathbb{C}^{K\times N}$ and $\mathbf{I}$} & \textcolor{blue}{K \texttimes N matrix, and the identity matrix}\tabularnewline
\hline 
\textcolor{blue}{$\mathbf{W}$} & \textcolor{blue}{Precoding matrix}\tabularnewline
\hline 
\textcolor{blue}{$K$} & \textcolor{blue}{Average number of the users in area $A$}\tabularnewline
\hline 
\textcolor{blue}{$N_{t}$ } & \textcolor{blue}{Number of transmit antennas}\tabularnewline
\hline 
\textcolor{blue}{$N_{r}$ } & \textcolor{blue}{Number of receive antennas}\tabularnewline
\hline 
\textcolor{blue}{$d_{k}$} & \textcolor{blue}{Distance between the BS and the $k^{th}$ user}\tabularnewline
\hline 
\textcolor{blue}{$\alpha$} & \textcolor{blue}{Path loss exponent}\tabularnewline
\hline 
$\lambda_{b}$, $\lambda_{u}$ and $\lambda_{r}$ & Densities of BSs, users and targets\tabularnewline
\hline 
$EE$ & Energy efficiency\tabularnewline
\hline 
$P$ & Transmit power\tabularnewline
\hline 
\end{tabular}
\par\end{centering}

\textcolor{blue}{\protect\caption{\label{tab:1}Summary of Symbols and Notations.}
}
\end{table}

\section{System Model\label{sec:System-Model}}

\textcolor{blue}{We consider a downlink  DFRC network that is designed
to serve randomly positioned communication users while at the same
time providing sensing services by detecting a number of randomly
positioned targets. The locations of DFRC BSs are modeled using Poisson
point process (PPP) $\Phi_{b}$ with density $\lambda_{b}$. The users
are also generated from an independent PPP $\Phi_{u}$ with density
$\lambda_{u}$, where each user is equipped with a single antenna.
Number of the users in area $A$ has a Poisson distribution with mean
$K=\lambda_{u}A$. In addition, the targets density is $\lambda_{r}$,
and each target flies at an altitude of $h_{t}$ \cite{Ghaith}. The
DFRC BSs are equipped with multiple antennas with $N_{t}$ transmit
and $N_{r}$ receive antennas and each user has a single antenna,
without loss of generality, it is assumed that $N_{t}<N_{r}$. For
clarity, Table \ref{tab:1} summarizes the commonly used symbols and
notations. Following \cite{assumption1,assumption2,muplink}, we rely
on the following standard assumptions:}

\textcolor{blue}{1) All the BSs are assumed to be active and share
the same transmission bandwidth. }

\textcolor{blue}{2) It is assumed that the intensity of the users
is high enough ($\lambda_{u}\gg\lambda_{b}$) such that each BS will
have at least one user served per channel \cite{muplink}, and each
user is associated with the nearest BS. Thus, the BSs are fully loaded
with full queues, and thus the BSs always have data to transmit. The
analysis with partially loaded BSs has been postponed to future work. }

\textcolor{blue}{3) The channel between the BSs and the users are
Rayleigh flat fading and perfectly estimated at the BSs. }

\textcolor{blue}{4) The DFRC system employs a DFRC waveform that is
both a communication signal and a radar probing waveform \cite{survy}. }

We leverage recent results in \cite{Spectral} that characterize the
radar mutual information, to evaluate the radar information rate performance.
A relevant performance metric for the design of ISAC networks is the
potential spectral efficiency (PSE), which is the network information
rate per unit area (measured in $bit/sec/m^{2}$) that corresponds
to the minimum signal quality for reliable transmission. 

In this work, we study DFRC  network deployment from a network-level
perspective. The main aim is to optimize the density of BSs to maximize
the energy efficiency of the network. The energy efficiency (EE) is
defined as a function of the communications and sensing rates and given
by 
\vspace{-2mm}
\begin{equation}
EE=\frac{R_{c}+R_{r}}{P_{T}}\label{eq:1},
\vspace{-2mm}
\end{equation}

\noindent where $R_{c}$ is the PSE of communication,
$R_{r}$ is the PSE of the radar system, and
$P_{T}$ is the total consumed power. Next, analytical expressions
of PSE for communication, $R_{c}$, and radar, $R_{r}$, are derived.

\section{PSE of DFRC system\label{sec:PSE-of-RadCom}}

\textcolor{blue}{Due to the stationarity and the independence of all the points in any homogeneous PPP, the performance of the typical user and typical target have the same characteristics as those of any other users and targets in the networks. Therefore, we analyze the performance of typical user/targets to represent the average performance of communication and sensing in the ISAC networks. Without loss of generality, the typical user/target is located at the origin and served by its closest BS.}

As we mentioned earlier, PSE is the network information rate per unit
area which can be calculated in general form by \cite{muplink,Pelaez,avgrate,book,Marco3}
\vspace{-2mm}
\begin{equation}
R=\lambda_{b}\log_{2}\left(1+\gamma\right)\textrm{Pr}\left(\gamma_{0}>\gamma\right),
\vspace{-2mm}
\end{equation}

\noindent \textcolor{blue}{where $\gamma_{0}$ is the received signal
to interference and noise ratio (SINR) at the typical user in communication
systems and at the BS in radar systems} and $\gamma$ is the SINR threshold
for reliable decoding. We note that, a fixed rate per user is defined,
i.e., $\log_{2}\left(1+\gamma\right)$ bits/sec/Hz, and the user is
served as long as the SINR allows it. This applies to delay limited
transmission where the spectral efficiency is determined by evaluating
the outage probability at a fixed rate, thus the outage probability
plays a pivotal role.

\subsection{PSE of communication system}

In the communication system the analysis applies to a typical user, as
permissible in any homogeneous PPP according to the Slivnyak\textendash Mecke\textquoteright s
theorem%
\footnote{Due to the independence between the points in PPP, conditioning on
a point at x does not change the distribution of the rest of the process.
This is very important theorem/property, and can be applied to any
user/target in the network.%
} \cite{book}. \textcolor{blue}{The typical user, denoted by $\textrm{\ensuremath{U_{o}}}$,}
and located at the origin. The PSE in this case can be obtained by
computing the PSE of $\textrm{\ensuremath{U_{o}}}$ and then averaging
the obtained conditional PSE with respect to all possible realizations
for the locations of the BSs and users. The PSE of the communication system
can be calculated by \cite{Marco}
\vspace{-2mm}
\begin{equation}
R_{c}=\lambda_{b}\log_{2}\left(1+\gamma_{c}\right)\textrm{Pr}\left(\gamma_{o}>\gamma_{c}\right)\label{eq:3},
\end{equation}
\noindent where $\gamma_{c}$ is the threshold for reliable decoding
of the communication message and $\gamma_{o}$ is the receive SINR
at the typical user $\textrm{\ensuremath{U_{o}}}$. Universal frequency
reuse is assumed, and thus each user not only receives information
from its home BS, but also suffers interference from all the other
BSs. The typical user $\textrm{\ensuremath{U_{o}}}$ is associated
with the nearest BS $b$. The signal transmitted by BS $b$ is denoted by $\mathbf{s}_{b}$. Thus, the received signal for
a typical user can be written as \cite{Marco,Marco3,Ghaith}
\vspace{-2mm}
\begin{equation}
\begin{aligned}
y_{o}=&\sqrt{P}d_{b,o}^{-\frac{\alpha}{2}}\mathbf{h}_{b,o}^{\dagger}\mathbf{w}_{b,o}s_{b,o}+
\underbrace{\sqrt{P}d_{b,o}^{-\frac{\alpha}{2}}\underset{k\in\Phi_{u}\setminus o}{\sum}\mathbf{h}_{b,o}^{\dagger}\mathbf{w}_{b,k}s_{b,k}}_{\text{BS } b {\text{'s signals to other users}}} \\
&+\underbrace{\underset{l\in\Phi_{b}\setminus b}{\sum}\sqrt{P}d_{l,o}^{-\frac{\alpha}{2}}\mathbf{h}_{l,o}^{\dagger}\mathbf{W}_{l}\mathbf{s}_{l}}_{{\text{Signals transmitted from other BSs}}}+n_{o},
\end{aligned}
\end{equation}
\noindent \textcolor{blue}{where $l$ and $k\in\mathbb{R}^{2}$},
$\mathbf{h}_{i,o}$ is an $N_{t}\times1$ vector denoting the small
scale fading between the $i^{th}$ BS and the $\textrm{M\ensuremath{U_{o}}}$
user, $d_{b,o}$ is the distance between the serving BS $b$ and
$\textrm{M\ensuremath{U_{o}}}$ user, while $d_{i,0}$ is the distance
from the $i^{th}$ BS and the $\textrm{M\ensuremath{U_{o}}}$ user.
The path loss exponent is $\alpha$, the precoding vector at BS $b$
of user $k$ is $\mathbf{w}_{b,k}$, the total precoding matrix at
BS $l$ is $\mathbf{W}_{l}$, $P$ is the transmit power, and $n_{o}$
denotes the additive white Gaussian noise (AWGN) at the user, i.e.,
$n_{o}\sim\mathcal{CN}\left(\text{0, }\sigma_{o}^{2}\right)$. Therefore,
the received SINR at the typical user $\textrm{\ensuremath{U_{o}}}$ is
given by 
\vspace{-2mm}
\begin{equation}
\gamma_{o} \!=\!\frac{Pd_{b,o}^{-\alpha}\left|\mathbf{h}_{b,o}^{\dagger}\mathbf{w}_{b,o}\right|^{2}}{Pd_{b,o}^{-\alpha}\underset{k\in\Phi_{u}\setminus o}{\sum}\left|\mathbf{h}_{b,o}^{\dagger}\mathbf{w}_{b,k}\right|^{2}+P\underset{l\in\Phi_{b}\setminus b}{\sum}d_{l,o}^{-\alpha}\left\Vert \mathbf{h}_{l,o}^{\dagger}\mathbf{W}_{l}\right\Vert ^{2} \!+ \! \sigma_{o}^{2}}\label{eq:7}.
\vspace{-2mm}
\end{equation}
Applying zero-forcing (ZF) precoding, the pseudo-inverse of the downlink channel
at the BS $b$ is, 
\begin{equation}
\mathbf{F}_{b}=\,\mathbf{H}_{b}^{H}\left(\mathbf{\mathbf{H}}_{b}\mathbf{H}_{b}^{H}\right)^{-1}=\left[\mathbf{f}_{b,k}\right]_{1\leq k\leq u}\label{eq:24-1},
\end{equation}
\noindent \textcolor{blue}{where $\mathbf{\mathbf{H}}_{b}$ is the
channel matrix between the }BS $b$\textcolor{blue}{{} and the users,
$\left[\mathbf{f}_{b,k}\right]_{k}$ is the kth column in the matrix
$\mathbf{F}_{b}$.} The precoding vector at BS $b$ for the typical
user can be written as $\mathbf{w}_{b,o}=\frac{\mathbf{f}_{b,o}}{\left\Vert \mathbf{f}_{b,o}\right\Vert }$.
Thus, the SINR expression in (\ref{eq:7}) can be written as 
\begin{equation}
\gamma_{o}=\frac{Pd_{o,o}^{-\alpha}\varsigma}{P\underset{l\in\Phi_{b}\setminus b}{\sum}d_{l,o}^{-\alpha}\left\Vert \mathbf{h}_{l,o}^{\dagger}\mathbf{W}_{l}\right\Vert ^{2}+\sigma_{o}^{2}}\label{eq:8},
\end{equation}
\noindent where $\varsigma=\frac{1}{\left\Vert \mathbf{f}_{b,o}\right\Vert ^{2}}=\frac{1}{\left[\left(\mathbf{\mathbf{H}}\mathbf{H}^{H}\right)^{-1}\right]_{o,o}}$.
Using (\ref{eq:3}) and (\ref{eq:8}) the PSE of the communication
system can be calculated by the expression presented in the next Theorem.
\begin{thm}
\noindent \textcolor{blue}{The PSE of the communication system can be
evaluated by (\ref{LongEquation1}), as shown at the top of this page,}
\noindent \textcolor{blue}{where $r_{i}$ and $\textrm{H}_{i}$ are
the $i^{th}$ zero and the weighting factor of the Laguerre polynomials,
respectively, and the remainder $R_{i}$ is negligible for Laguerre
polynomials order $n>15$ \cite{book2} and $\kappa=\frac{K}{N}.$}\end{thm}
\begin{IEEEproof}
The proof is provided in Appendix A.
\end{IEEEproof}
\noindent 

\begin{figure*}
	\begin{equation}\label{LongEquation1}
		\begin{aligned}
			R_{c} \!=& \!\lambda_{b}\log_{2}\left(1+\gamma_{c}\right)2\pi\lambda_{b}\stackrel[i=1]{n}{\sum}\textrm{H}_{i}e^{r_{i}}r_{i} \\
			&\times\stackrel[k=0]{\kappa-1}{\Pi}e^{-\left(\frac{2\pi\lambda_{b}\left(\kappa-1\right)!}{\Gamma\left(\kappa\right)}\left(\frac{1}{\alpha\, k!}\left(\left(\beta\left(\left(\frac{r_{i}^{-\alpha}}{\gamma_{c}}-\frac{\sigma_{0}^{2}}{P}\right)\right)\right)^{\frac{-2}{\alpha}}\right)\Gamma\left(\frac{k+2}{\alpha},\beta\left(\left(\frac{r_{i}^{-\alpha}}{\gamma_{c}}-\frac{\sigma_{0}^{2}}{P}\right)\right)\right)\right)+\frac{\pi\lambda_{b}r_{i}^{2}}{\kappa}\right)}\!+\!R_{i},
		\end{aligned}
	\end{equation}
\end{figure*}

In this work, ZF precoding is applied at the BSs, and the interference
from other BSs at the typical user is considered under the assumption
of an MU-MIMO network operation with single-antenna uses. In Appendix
A, we present different approaches to derive exact and approximation
expressions for the PSE in such a communication network. Firstly,
we show that the exact expression can be obtained by deriving the
cumulative distribution function (CDF) of the aggregated interference term. Thus, we first derive the
moment generating function (MGF) of the aggregated interference term, then using the inverse Laplace
transform or the Gil\textendash Pelaez inversion theorem, we obtain
the CDF of the aggregated interference term. However, the exact expression
presented in \textcolor{blue}{(\ref{eq:91})} is complex, and does
not present a tractable form. In order to obtain a simpler closed-form
expression, we derive a tighter approximation, by only considering
the dominant interferences. This technique has been considered in
the literature, because of its simplicity and accuracy. We would like
to indicate that the novelty in terms of analytical derivations in
this work comes from using the joint comms-radar rate, as will be
presented in the next sections.

\subsection{PSE of radar system}

In radar systems, without any loss in generality, the analysis is conducted
on a tagged BS, the $b$th BS, as the reference. The analysis holds
for a generic BS located at a generic location \cite{muplink}. The
PSE of radar systems can be calculated by 
\begin{equation}
R_{r}=\lambda_{b}\log_{2}\left(1+\gamma_{r}\right)\textrm{Pr}\left(\gamma_{b}>\gamma_{r}\right)\label{eq:10},
\end{equation}
\noindent where $\gamma_{r}$ is the threshold for reliable decoding
and $\gamma_{b}$ is the receive SINR at the $b$th BS. 

\textcolor{blue}{Note that, the rate in radar systems has been defined
in the literature. Most radar systems operate by radiating an electromagnetic
signal into a region and detecting the echo returned from the reflecting
targets. The nature of the echo signal provides information about
the target, such as range, radial velocity, angular direction, size,
shape, and so on. This signal is usually referred to as the radar
waveform, and plays a key role in the accuracy, resolution, and ambiguity
of radar in performing the above-mentioned tasks. In fact, the application
of information theory to radar can be traced to the early 1950s when
Woodward and Davies examined the use of information-theoretic principles
to obtain the a posteriori radar receiver, shortly after the publication
of Shannon\textquoteright s milestone work in information theory.
Woodward and Davies give an excellent example of how one can use information
theory to benefit radar system design \cite{woodward1950xcii}. Many researchers considered
the connection between information theory and radar design problems
until in 1993, Bell published his paper that suggested maximizing
the mutual information between the target impulse response and the
reflected radar signal to design radar waveforms. It was noticed that
it is implied that the greater rate between the target impulse response
(the target reflection) and the reflected signals, the better capability
of radar to estimate the parameters describing the target. Radar
rate is an appropriate metric to characterize the estimation accuracy
of the system parameters, thus the more radar rate the better performance
we can achieve }\cite{radarrate1,radarrate2,Spectral}\textcolor{blue}{.}

Consider the typical target, $b\in\Phi_{r}$, the received
signal at the served BS, i.e., $b$th BS, can be written as
\begin{equation}
\mathbf{y}_{b}=\sqrt{P}\mathbf{G}_{b}\mathbf{W}_{b}\mathbf{s}_{b}+\underset{\textrm{Interference from DL BSs}}{\underbrace{\underset{l\in\Phi_{b}\setminus b}{\sum}\sqrt{P}d_{l,b}^{-\frac{\alpha}{2}}\mathbf{H}_{l,b}\mathbf{W}_{l}\mathbf{s}_{l}}}+\mathbf{n}_{b},
\end{equation}
\noindent where $\mathbf{G}_{b}$ is an $N_{r}\times N_{t}$ target
response matrix at the $b$th BS and given by, $\mathbf{G}_{b}=\alpha_{b}\mathbf{a}\left(\theta_{b}\right)\mathbf{b}^{T}\left(\theta_{b}\right)$,
$\alpha_{b}$ is the amplitude of the $b$th target which contains
the round-trip path-loss and the radar cross-section of the target,
$\mathbf{a}\left(\theta_{b}\right)\textrm{and }\mathbf{b}^{T}\left(\theta_{b}\right)$
are the associated transmit and receive array steering vectors, respectively,
and $\theta_{b}$ is its direction of arrival (DOA), $\mathbf{H}_{l,b}^{\dagger}$
is the $N_{r}\times N_{t}$ channel matrix between the $l^{th}$ BS
and $b$th BS, $\mathbf{W}_{i}$ is the $N_{t}\times K$ precoding
matrix at BS $i$, and $\mathbf{n}_{b}$ is the disturbance in the
receiver (accounting for the internal thermal noise, the sky noise,
external disturbance, clutter, etc.) \cite{Spectral}. Please note
that the above modeling implies an inherent assumption that the targets
are line-of-sight (LoS), otherwise they are non-detectable.\textcolor{blue}{{}
For simplicity, the reflection terms from the other targets in the system are assumed to
be small after the receive filtering operation and thus it can be ignored
without any major impact on the system design. }The radar receiver
uses a filter $\mathbf{w}_{r}$ to reduce the interference and noise.
Then, the filtered signal can be written by
\begin{equation}
\begin{aligned}
	\mathbf{w}_{r}^{H}\mathbf{y}_{b}=&\sqrt{P}\mathbf{w}_{r}^{H}\mathbf{G}_{b}\mathbf{W}_{b}\mathbf{s}_{b}\\
	&+\underset{l\in\Phi_{b}\setminus b}{\sum}\sqrt{P}d_{l,b}^{-\frac{\alpha}{2}}\mathbf{w}_{r}^{H}\mathbf{H}_{l,b}\mathbf{W}_{l}\mathbf{s}_{l}+\mathbf{w}_{r}^{H}\mathbf{n}_{b}\label{eq:20},
\end{aligned}
\end{equation}
\noindent where $\mathbf{w}_{r}=\frac{\mathbf{R^{-1}}\,\mathbf{a}}{\mathbf{a}^{H}\,\mathbf{R}^{-1}\,\mathbf{a}}$,
is the minimum variance distortion-less response (MVDR) beamformer,
and $\mathbf{R}=\mathscr{E}\left\{ \mathbf{\tilde{n}}\mathbf{\tilde{n}}^{H}\right\} =\left\{ \underset{l\in\Phi_{b}\setminus b}{\sum}\underset{k\in\Phi_{u}}{\sum}Pd_{l,b}^{-\alpha}\mathbf{z}\mathbf{z}^{H}+\mathbf{I}\sigma_{b}^{2}\right\} $ and
$\mathbf{z}=\mathbf{H}_{l,b}\mathbf{w}_{l,k}$ \cite{mvdr1}. The
expression in (\ref{eq:20}) can be simplified to
\begin{equation}
\tilde{y}_{b}=\sqrt{P}\mathbf{w}_{r}^{H}\mathbf{G}_{b}\mathbf{W}_{b}\mathbf{s}_{b}+\underset{l\in\Phi_{b}\setminus b}{\sum}\sqrt{P}d_{l,b}^{-\frac{\alpha}{2}}\mathbf{w}_{r}^{H}\mathbf{H}_{l,b}\mathbf{W}_{l}\mathbf{s}_{l}+\mathbf{w}_{r}^{H}\mathbf{n}_{b}.
\end{equation}
Therefore, the output SINR of the radar system is 
\begin{equation}
\gamma_{b}=\frac{\left|\alpha_{b}\right|^{2}P\mathbf{w}_{r}^{H}\mathbf{G}_{b}\mathbf{W}_{b}\mathbf{s}_{b}\mathbf{s}_{b}^{H}\mathbf{W}_{b}^{H}\mathbf{G}_{b}^{H}\mathbf{w}_{r}}{\underset{l\in\Phi_{b}\setminus b}{\sum}P\mathbf{w}_{r}^{H}\mathbf{H}_{l,b}\mathbf{W}_{l}\mathbf{s}_{l}\mathbf{s}_{l}^{H}\mathbf{W}_{l}^{H}\mathbf{H}_{l,b}^{H}\mathbf{w}_{r}+\left\Vert \mathbf{w}_{r}^{H}\right\Vert ^{2}\sigma_{b}^{2}}\label{eq:17}.
\end{equation}

\noindent Using (\ref{eq:10}) and (\ref{eq:17}) the PSE of the radar
system can be calculated by the expression presented in the next Theorem.
\begin{thm}
\noindent The PSE of the radar system can be evaluated by
\[
R_{r}=\lambda_{b}\log_{2}\left(1+\gamma_{r}\right)\left(1-\stackrel[i=0]{N-1}{\sum}\frac{4\pi\lambda_{b}}{i!}\stackrel[n=1]{Q}{\sum}\textrm{H}_{n}e^{r_{n}}r_{n}\left(-\frac{\tilde{\alpha}\left(N-1\right)\left(\alpha-2\right)P\tilde{\alpha_{b}}r_{n}^{-\alpha_{r}}}{2\pi\lambda_{b}\beta r^{2-m}\gamma_{r}}\right)^{i}\right.
\]

\begin{equation}
	\left.\times e^{-\left(\frac{\left(N-1\right)\left(\alpha-2\right)\alpha P\tilde{\alpha_{b}}r_{n}^{-\alpha_{r}}}{2\pi\lambda_{b}\beta r^{2-\alpha}\gamma_{r}}+2\pi\lambda_{b}\left(r_{n}^{2}-h_{t}^{2}\right)\right)}+R_{n}\right)
\end{equation}

\noindent where $r_{r_{n}}$ and $\textrm{H}_{n}$ are the $n^{th}$
zero and the weighting factor of the Laguerre polynomials, respectively,
and the remainder $R_{n}$ is negligible for $Q>15$ \cite{book2}.\end{thm}
\begin{IEEEproof}
The proof is provided in Appendix B in \cite{salem2022rethinking} due to page limitation.

Finally, the $EE$ can be evaluated by substituting $R_{c}$ in Theorem
1 and $R_{r}$ in Theorem 2 into (\ref{eq:1}). 
\end{IEEEproof}

\section{Optimal Density of BSs \label{sec:Optimal-BS-Density}}

In this section, we analyze whether there is an optimal and unique density of BSs in DFRC, communications-only and radar-only networks.

\subsection{ISAC Network}

In this sub-section, we consider whether there is an optimal and unique
density of DFRC BSs that maximizes the energy efficiency while all
the other system parameters are fixed and given. Mathematically, the
optimization problem can be formulated as 
\[
\underset{\lambda_{b}}{\max}\:\frac{R_{c}+R_{r}}{P_{T}}
\]
\begin{equation}
\textrm{subject to}\:\lambda_{b}\in\left[\lambda_{b}^{\min},\lambda_{b}^{\max}\right]\label{eq:18}.
\end{equation}

\noindent Following \cite{Marco,powermodel} we define $P_{T}=\lambda_{b}\left(P_{tx}+P_{circ}\right)$
as the network power consumption which can be obtained by multiplying
the average number of BSs per unit area, i.e., $\lambda_{b}$, and
the average power consumption of a BS, which is $P_{tx}+P_{circ}$
where $P_{tx}=\frac{\bar{P}_{tx}}{\eta_{eff}}$, $\bar{P}_{tx}$ is
the power consumption due to the transmit power, $\eta_{eff}$ is
the efficiency of the amplifier and the antennas, $P_{circ}$
the static (circuit) power, and $\lambda_{b}^{\min},\lambda_{b}^{\max}$
are the minimum and maximum allowed density of the BSs, respectively.
Without loss of generality, we can assume: $\lambda_{b}^{\min}\rightarrow0,\lambda_{b}^{\max}\rightarrow\infty$.
By substituting $R_{c}$ in Theorem 1 and $R_{r}$ in Theorem 2 in
(\ref{eq:18}), and considering first-order Laguerre polynomial the
optimization problem can be written in a more detailed formula according to the derived closed-form expression of communication rate and radar rate. Then, the formulated problem can be simplified to 
\[
\underset{\lambda_{b}}{\max}\: \tilde EE
\]
\begin{equation}
\textrm{subject to}\:\lambda_{b}\in\left[\lambda_{b}^{\min},\lambda_{b}^{\max}\right]\label{eq:21},
\end{equation}
\noindent where $\tilde EE = \left(a_{1}\lambda_{b}\stackrel[k=0]{\kappa-1}{\Pi}e^{-\lambda_{b}a_{2,\kappa}}\right)/{\left(P_{tx}+P_{circ}\right)}+b_{5}\left(1-\lambda_{b}e^{-\left(\frac{b_{3}}{\lambda_{b}}+b_{4}\lambda_{b}\right)}\stackrel[i=0]{N-1}{\sum}b_{1,i}\left(-\frac{1}{\lambda_{b}}\right)^{i}b_{2,i}\right)/{\left(P_{tx}+P_{circ}\right)}$, $a_{1}=\log_{2}\left(1+\gamma_{c}\right)2\pi\textrm{H}_{1}e^{r_{1}}r_{1}$,
$a_{2,\kappa}=\frac{2\pi\left(\kappa-1\right)!}{\Gamma\left(\kappa\right)}\left(\frac{1}{\alpha\, k!}\left(\beta\left(\frac{r_{1}^{-\alpha}}{\gamma_{c}}\right)\right)^{\frac{-2}{m}}\Gamma\left(\frac{k+2}{\alpha},\beta\left(\frac{r_{1}^{-\alpha}}{\gamma_{c}}\right)\right)\right)+\frac{\pi r_{1}^{2}}{\kappa}$,
$b_{5}=\log_{2}\left(1+\gamma_{r}\right)$, $b_{1,i}=\frac{4\pi}{i!}\textrm{H}_{1}e^{r_{r_{1}}}r_{r_{1}}$,
$b_{2,i}=\left(-\frac{\tilde{\alpha}\left(N-1\right)\left(\alpha-2\right)P\tilde{\alpha_{b}}r_{r_{1}}^{-\alpha_{r}}}{2\pi\beta r^{2-m}\gamma_{r}}\right)^{i}$,
$b_{3}=\frac{\left(N-1\right)\left(\alpha-2\right)\tilde{\alpha}P\tilde{\alpha_{b}}r_{r_{1}}^{-\alpha_{r}}}{2\pi\beta r^{2-\alpha}\gamma_{r}}$,
and $b_{4}=2\pi\lambda_{b}\left(r_{r_{1}}^{2}-h_{t}^{2}\right)$.
Using the identity that $\stackrel[k=0]{\kappa-1}{\Pi}e^{-\lambda_{b}a_{2,\kappa}}=e^{-\lambda_{b}\stackrel[k=0]{\kappa-1}{\sum}a_{2,\kappa}}$,
the last expression in (\ref{eq:21}) can be written as 
\[
\underset{\lambda_{b}}{\max}\:\frac{a_{1}\lambda_{b}e^{-\lambda_{b}a_{3}} \!+\! b_{5}\left(1\!-\!\lambda_{b}e^{-\left(\frac{b_{3}}{\lambda_{b}}+b_{4}\lambda_{b}\right)}\stackrel[i=0]{N-1}{\sum}b_{1,i}(-\frac{1}{\lambda_{b}})^{i}b_{2,i}\right)}{\left(P_{tx}+P_{circ}\!\right)}
\]
\begin{equation}
\textrm{subject to}\:\lambda_{b}\in\left[\lambda_{b}^{\min},\lambda_{b}^{\max}\right],
\end{equation}

\noindent where $a_{3}=\stackrel[k=0]{\kappa-1}{\sum}a_{2,\kappa}$. 

With the aid of some algebraic manipulations, we can observe that,
the energy efficiency expression, i.e., the objective function, is
a uni-modal and pseudo-concave function in $\lambda_{b}$, which can be proved in a similar approach as demonstrated in reference \cite{Marco}. The optimal
value of $\lambda_{b}$ can be obtained as the unique solution of
the following equation

\begin{equation}
\frac{\partial EE}{\partial\lambda_{b}}=\frac{\partial EE_{c}}{\partial\lambda_{b}}+\frac{\partial EE_{r}}{\partial\lambda_{b}}=0\label{eq:26}.
\end{equation}

The derivative of the energy efficiency of communication system, $EE_{c}=\frac{R_{c}}{P_{T}}$,
with respect to $\lambda_{b}$ is given by

\begin{equation}
\frac{\partial EE_{c}}{\partial\lambda_{b}}=a_{1}e^{-a_{3}\lambda_{b}}-a_{1}a_{3}\lambda_{b}e^{-a_{3}\lambda_{b}}\label{eq:24}.
\end{equation}

Similarly, the derivative of the energy efficiency of radar system,
$EE_{r}=\frac{R_{r}}{P_{T}}$, with respect to $\lambda_{b}$ is given
by
\begin{align}
	&\frac{\partial EE_{r}}{\partial\lambda_{b}}=\frac{-\left(b_{1,i}b_{2,i}e^{-\left(\frac{b_{3}}{\lambda_{b}}+b_{4}\lambda_{b}\right)}\lambda_{b}\left(1+\left(-\left(\frac{1}{\lambda_{b}}\right)\right)^{N}\lambda_{b}\right)\right)}{\left(1+\lambda_{b}\right)^{2}} \nonumber \\
	&+\frac{\left(b_{1,i}b_{2,i}e^{-\left(\frac{b_{3}}{\lambda_{b}}+b_{4}\lambda_{b}\right)}\left(1+\left(-\left(\frac{1}{\lambda_{b}}\right)\right)^{N}\lambda_{b}\right)\right)}{\left(1+\lambda_{b}\right)} \nonumber \\
	&+\!\frac{\left(b_{1,i}b_{2,i}e^{-\left(\frac{b_{3}}{\lambda_{b}}+b_{4}\lambda_{b}\right)}\lambda_{b}\left(1+\left(-\left(\frac{1}{\lambda_{b}}\right)\right)^{N}\lambda_{b}\right)\left(\frac{b_{3}-b_{4}}{\lambda_{b}^{2}}\right)\right)}{\left(1+\lambda_{b}\right)} \nonumber \\
	& +\!\frac{\left(b_{1,i}b_{2,i}e^{-\left(\frac{b_{3}}{\lambda_{b}}+b_{4}\lambda_{b}\right)}\lambda_{b}\left(\left(-\frac{1}{\lambda_{b}}\right)^{N}+\frac{\left(-\frac{1}{\lambda_{b}}\right)^{-1+N}N}{\lambda_{b}}\right)\right)}{\left(1+\lambda_{b}\right)}\label{eq:25}.
\end{align}

However, the expressions in (\ref{eq:24}) and (\ref{eq:25}) are
complicated, and this makes (\ref{eq:26}) hard to solve. Nevertheless,
finding solutions to polynomial formulas is quite easy using numerical
methods, such as Newton's method. Newton's method can be explained
as follows.

\textcolor{blue}{\emph{Let $f$ be a differentiable function, and
the derivative of $f$ is $f'$. We seek a solution
of $f(x)=0$, starting from an initial estimate $x=x_{1}$. At the
$n$th step, given $x_{n}$, compute the next approximation $x_{n+1}$
by}}

\emph{$x_{n+1}=x_{n}-\frac{f(x_{n})}{f'(xn)}$and
repeat.}

In order to compare the optimal value of BSs density of DFRC system
with only communication system and only radar system, in the next
subsections we derive also the optimal $\lambda_{b}$ of only communication
and radar systems.

\subsection{Communications-only network}

Firstly, the optimal $\lambda_{b}$ of only communication system can
be found by solving the equation
\begin{equation}
\frac{\partial EE_{c}}{\partial\lambda_{b}}=a_{1}e^{-a_{3}\lambda_{b}}-a_{1}a_{3}\lambda_{b}e^{-a_{3}\lambda_{b}}=0,
\end{equation}
and 
\begin{equation}
e^{-a_{3}\lambda_{b}}=a_{3}\lambda_{b}e^{-a_{3}\lambda_{b}}.
\end{equation}
Thus the optimal $\lambda_{b}$ of only communication system is given
by
\begin{equation}
\lambda_{b}=\frac{1}{a_{3}}.
\end{equation}

\subsection{Radar-only network}
Secondly, the optimal $\lambda_{b}$ of only radar system can be obtained
by solving the equation
\begin{align}
	&\frac{\partial EE_{r}}{\partial\lambda_{b}}=\frac{-\left(b_{1,i}b_{2,i}e^{-\left(\frac{b_{3}}{\lambda_{b}}+b_{4}\lambda_{b}\right)}\lambda_{b}\left(1+\left(-\left(\frac{1}{\lambda_{b}}\right)\right)^{N}\lambda_{b}\right)\right)}{\left(1+\lambda_{b}\right)^{2}} \nonumber \\ 
	& +\frac{\left(b_{1,i}b_{2,i}e^{-\left(\frac{b_{3}}{\lambda_{b}}+b_{4}\lambda_{b}\right)}\left(1+\left(-\left(\frac{1}{\lambda_{b}}\right)\right)^{N}\lambda_{b}\right)\right)}{\left(1+\lambda_{b}\right)} \nonumber \\
	&+\frac{\left(b_{1,i}b_{2,i}e^{-\left(\frac{b_{3}}{\lambda_{b}}+b_{4}\lambda_{b}\right)}\lambda_{b}\left(1+\left(-\frac{1}{\lambda_{b}}\right)^{N}\lambda_{b}\right)\left(\frac{b_{3}}{\lambda_{b}^{2}}-b_{4}\right)\right)}{\left(1+\lambda_{b}\right)} \nonumber \\
	&+\frac{\left(b_{1,i}b_{2,i}e^{-\left(\frac{b_{3}}{\lambda_{b}}+b_{4}\lambda_{b}\right)}\lambda_{b}\left(\left(-\frac{1}{\lambda_{b}}\right)^{N}+\frac{\left(-\frac{1}{\lambda_{b}}\right)^{-1+N}N}{\lambda_{b}}\right)\right)}{\left(1+\lambda_{b}\right)} \nonumber \\
	&=0.
\end{align}

With the aid of some algebraic manipulations, we can get
\begin{equation}
\begin{aligned}
	\frac{\left(-1\right)^{N}\lambda_{b}^{2}\frac{1}{\lambda_{b}^{N}}\left(1-N+\lambda_{b}\left(1-N\right)\right)}{\left(1+\left(-1\right)^{N}\frac{1}{\lambda_{b}^{N}}\lambda_{b}\right)}\\-\lambda_{b}^{3}b_{4}-\lambda_{b}^{2}b_{4}+\lambda_{b}\left(b_{3}+1\right)=-b_{3}\label{eq:31}.
\end{aligned}
\end{equation}
In case $\lambda_{b}<1$ we can get \textcolor{blue}{$\left(1+\left(-1\right)^{N}\frac{1}{\lambda_{b}^{N}}\lambda_{b}\right)\simeq\left(-1\right)^{N}\frac{1}{\lambda_{b}^{N}}\lambda_{b}$.}
Therefore, (\ref{eq:31}) can be simplified to
\begin{equation}
\begin{aligned}
	\frac{\left(-1\right)^{N}\lambda_{b}^{2}\frac{1}{\lambda_{b}^{N}}\left(1-N+\lambda_{b}\left(1-N\right)\right)}{\left(-1\right)^{N}\frac{1}{\lambda_{b}^{N}}\lambda_{b}}\\  -\lambda_{b}^{3}b_{4}-\lambda_{b}^{2}b_{4}+\lambda_{b}\left(b_{3}+1\right)=-b_{3},
\end{aligned}
\end{equation}
which can be written as 
\begin{equation}
\lambda_{b}^{3}b_{4}-\lambda_{b}^{2}\left(1-N-b_{4}\right)-\lambda_{b}\left(1-N+b3+1\right)-b_{3}=0\label{eq:33},
\end{equation}
where $b_{3}=\frac{\left(N-1\right)\left(\alpha-2\right)\tilde{\alpha}P\tilde{\alpha_{b}}r_{r_{1}}^{-\alpha_{r}}}{2\pi\beta r^{2-\alpha}\gamma_{r}}$
and $b_{4}=2\pi\lambda_{b}\left(r_{r_{1}}^{2}-h_{t}^{2}\right)$. Last equation in (\ref{eq:33}) can be written in more general from
as
\begin{equation}
a\lambda_{b}^{3}-b\lambda_{b}^{2}-c\lambda_{b}-d=0,
\end{equation}
which has solution given by (\ref{eq:30}).

\begin{figure*}
	\begin{equation}\label{eq:30}
	\begin{aligned}	\lambda_{b}=\frac{b}{3a}-\frac{2^{\frac{1}{3}}\left(-b^{2}-3ac\right)}{3a\left(2b^{3}+9abc+27a^{2}d+\sqrt{4\left(-b^{2}-3ac\right)^{3}+\left(2b^{3}+9abc+27a^{2}d\right)^{2}}\right)^{\frac{1}{3}}} \\
		+\frac{\left(2b^{3}+9abc+27a^{2}d+\sqrt{4\left(-b^{2}-3ac\right)^{3} +\left(2b^{3}+9abc+27a^{2}d\right)^{2}}\right)^{\frac{1}{3}}}{3\,2^{\frac{1}{3}}a}.
	\end{aligned}
	\end{equation}
\end{figure*}

In case $\lambda_{b}>1$ we can get\textcolor{blue}{ $\left(1+\left(-1\right)^{N}\frac{1}{\lambda_{b}^{N}}\lambda_{b}\right)\backsimeq1.$}
Therefore, (\ref{eq:31}) can be simplified to
\begin{equation}
\begin{aligned}
	\left(-1\right)^{N}\left(\lambda_{b}^{2-N}-N\lambda_{b}^{2-N}+\lambda_{b}^{3-N}\left(1-N\right)\right) \\
	-\lambda_{b}^{3}b_{4}-\lambda_{b}^{2}b_{4}+\lambda_{b}\left(b_{3}+1\right)=-b_{3},
\end{aligned}
\end{equation}
and
\begin{equation}
\lambda_{b}^{3}b_{4}+\lambda_{b}^{2}b_{4}-\lambda_{b}\left(b_{3}+1\right)-b_{3}=0,
\end{equation}
which can be written in more general from
\begin{equation}
a\lambda_{b}^{3}+b\lambda_{b}^{2}-c\lambda_{b}-d=0\label{eq:33-2},
\end{equation}
which has solution given by (\ref{eq:34}).

\begin{figure*}
	\begin{equation}\label{eq:34}
	\begin{aligned}
		\lambda_{b}=\frac{b}{3a}-\frac{2^{\frac{1}{3}}\left(-b^{2}-3ac\right)}{3a\left(-2b^{3}-9abc+27a^{2}d+\sqrt{4\left(-b^{2}-3ac\right)^{3}+\left(-2b^{3}-9abc+27a^{2}d\right)^{2}}\right)^{\frac{1}{3}}} \\
		+\frac{\left(-2b^{3}-9abc+27a^{2}d+\sqrt{4\left(-b^{2}-3ac\right)^{3}+\left(-2b^{3}-9abc+27a^{2}d\right)^{2}}\right)^{\frac{1}{3}}}{3\,2^{\frac{1}{3}}a}.
	\end{aligned}
	\end{equation}
	
\end{figure*}

\textcolor{blue}{These solutions of the BS densities presented in
(\ref{eq:30}) and (\ref{eq:34}) depend on the system parameters.
If there are more than one real and positive solutions, the optimal BS density is determined by plugging the solutions yielding the highest EE into EE.}

\section{Numerical Results\label{sec:Numerical-Results}}

In this section, we present some numerical results in order to verify
the accuracy of the derived analytical framework, to calculate the
energy efficiency, and to show the results of the EE optimization
problem as a function of the BSs density. Unless otherwise stated,
the simulation setup is summarized in Table \ref{tab:Para}. \textcolor{blue}{Monte
Carlo simulations are executed by simulating several realizations,
according to the PPP model, of the ISAC network and by computing the
PSE and the power consumption. We have used the following methodology
as in \cite{Montecarlo}. Step 1 : circular area of radius $\textrm{\ensuremath{\left(R_{A}\right)}}$
around the origin is considered. Step 2 : number of BSs is generated
following a Poisson distribution. Step 3 : The BSs are uni-formally
distributed over the circular region. Step 4 : Independent channel
gains are generated for each BS, user, and target. Step 5 : the SINR
is computed. Step 6 : Finally, the EE is computed by repeating Step
1\textendash Step 5 for $10^{5}$ times, and then all the performance results of all snapshots are averaged.}

\begin{table}
\noindent \begin{centering}
\begin{tabular}{|c|c|}
\hline 
Parameter & Value\tabularnewline
\hline 
\hline 
\textcolor{blue}{Monte-Carlo simulations repeated} & \textcolor{blue}{$10^{5}$times}\tabularnewline
\hline 
\textcolor{blue}{Path loss exponent} & \textcolor{blue}{2.7}\tabularnewline
\hline 
\textcolor{blue}{The radius of the area $\textrm{\ensuremath{\left(R_{A}\right)}}$ } & \textcolor{blue}{250 m}\tabularnewline
\hline 
\textcolor{blue}{Bw} & \textcolor{blue}{20 MHz}\tabularnewline
\hline 
\textcolor{blue}{$\textrm{P}_{\textrm{circ}}$} & \textcolor{blue}{51.14 dBm}\tabularnewline
\hline 
\textcolor{blue}{$\bar{P}_{tx}$} & \textcolor{blue}{43 dBm}\tabularnewline
\hline 
\textcolor{blue}{$\eta_{eff}$} & \textcolor{blue}{0.5}\tabularnewline
\hline 
\textcolor{blue}{Number of transmit antennas } & \textcolor{blue}{4}\tabularnewline
\hline 
\textcolor{blue}{Number of receive antennas } & \textcolor{blue}{8}\tabularnewline
\hline 
\textcolor{blue}{The height of BS} & \textcolor{blue}{25 $m$}\tabularnewline
\hline 
\textcolor{blue}{The height of users} & \textcolor{blue}{1.5 $m$}\tabularnewline
\hline 
\textcolor{blue}{The height of street level target} & \textcolor{blue}{1.5 $m$}\tabularnewline
\hline 
\textcolor{blue}{The height of lower-level aerial target } & \textcolor{blue}{50 $m$}\tabularnewline
\hline 
\textcolor{blue}{The height of higher-level aerial target } & \textcolor{blue}{200 $m$}\tabularnewline
\hline 
\end{tabular}
\par\end{centering}
\protect\caption{\label{tab:Para}Simulation Parameters.}
\end{table}

\begin{figure*}
\noindent \begin{centering}
\subfloat[Height of the target
$h_{t}=1.5m$.]{\noindent \begin{centering}
\includegraphics[scale=0.5]{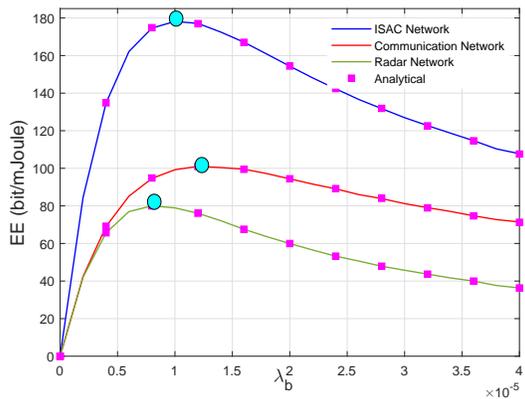}
\par\end{centering}
}\subfloat[Height of the target
$h_{t}=50m$.]{\noindent \begin{centering}
\includegraphics[scale=0.5]{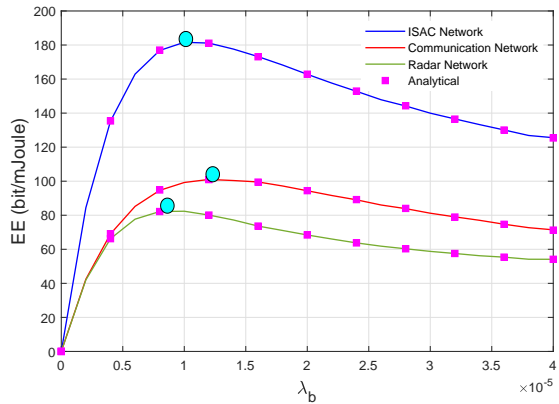}
\par\end{centering}
}
\par\end{centering}
\noindent \begin{centering}
\subfloat[Height of the target
$h_{t}=200m$.]{\noindent \begin{centering}
\includegraphics[scale=0.5]{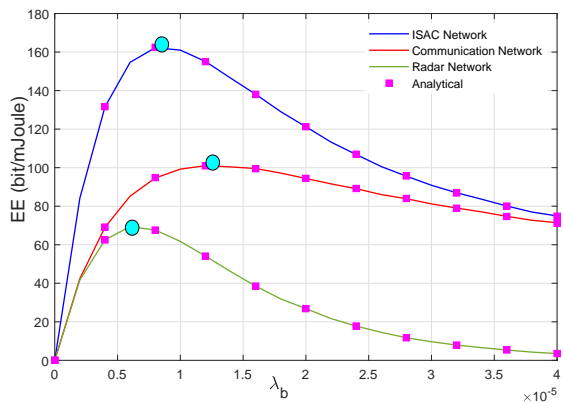}
\par\end{centering}
}
\par\end{centering}
\protect\caption{\label{fig:1}Energy efficiency versus the BS density for different
values of the target height, the circles represent the optimal BS
density. }
\end{figure*}

In Fig. \ref{fig:1}, we plot the energy efficiency versus the density
of the BSs for different target heights: street-level targets at $h_{t}=1.5m$,
lower-level aerial targets at $h_{t}=50m$, and higher-level aerial
targets at $h_{t}=200m$. The results depicted in Fig. \ref{fig:1}
confirm the good accuracy of the derived analytical approach. In addition,
these results show clearly \textcolor{blue}{the uni-modal and pseudo-concave
shape of the energy efficiency} as a function of the BSs\textquoteright{}
density for a given system parameters. Accordingly, there is an optimal
value of the BSs density that maximizes the system performance. 
 \textcolor{blue}{Importantly, the optimal BS density for ISAC networks is lower than that for classical communication-only networks. This is expected since the sensing component can effectively compensate for the potential efficiency loss in the communication systems.} However, interestingly
enough, this optimal value becomes smaller when the target altitude
increases. That means, in very high altitude cases few number of BSs
is adequate to reach the optimal performance due to weak interference
power. Moreover, the energy efficiency performance of communication
system is always better than the radar system which goes to zero as
the target flies higher than 200 m for a given system set up. This
is a result of the two-way propagation and hence the increased path
loss experienced by the radar signal propagation.

It is worth mentioning that based on the analysis in \cite{radius1,radius2,Marco},
the average cell radius can be computed as $\mathfrak{R}^{c}=\frac{1}{\sqrt{\pi\lambda_{b}}}$.
Accordingly, the optimal cell radius for communication network only
is $\mathfrak{R}^{c}=162.8675\textrm{ m}$, while the optimal cell
radius for ISAC network is $\mathfrak{R}^{c}=178.4124\textrm{ m}$
in street-level targets and lower-level aerial targets cases, and
$\mathfrak{R}^{c}=199.4711\textrm{ m}$ in higher-level aerial targets
case. {\textcolor{blue}{The main reason is that the sensing component within the ISAC networks can effectively compensate for the potential energy efficiency loss in the communication systems under a lower BS density. As a result, ISAC networks can attain heightened energy efficiency even with lower BS density, i.e., thereby bringing a larger equivalent coverage radius.}}

\begin{figure*}
\noindent \begin{centering}
\subfloat[\label{fig:2a}Energy efficiency versus the reliability thresholds
$\gamma_{c}$ and $\gamma_{r}$ \\ when $\gamma_{c}=\gamma_{r}=\gamma\textrm{ \ensuremath{\left(dB\right)}}$
and the height of the target $h_{t}=1.5\textrm{ m}.$]{\noindent \begin{centering}
\includegraphics[scale=0.55]{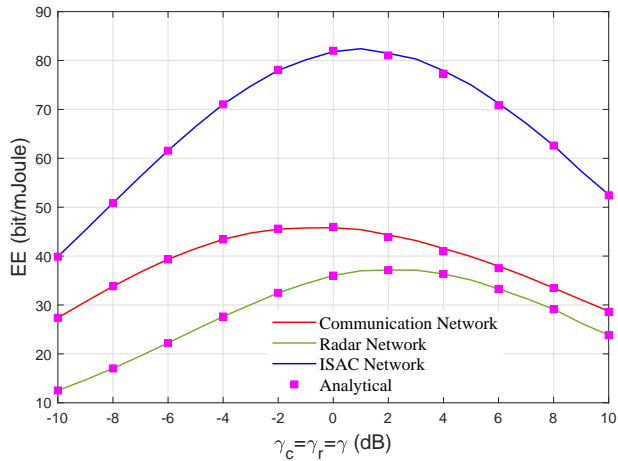}
\par\end{centering}

}\subfloat[\label{fig:2b}Energy efficiency versus the reliability thresholds
$\gamma_{c}$ and $\gamma_{r}$ when $\gamma_{c}=\gamma_{r}=\gamma\textrm{ \ensuremath{\left(dB\right)}}$
and the height of the target $h_{t}=50\textrm{ m}.$]{\noindent \begin{centering}
\includegraphics[scale=0.55]{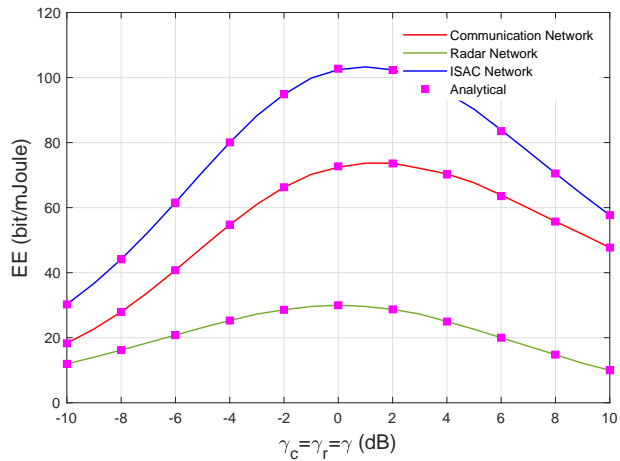}
\par\end{centering}

}
\par\end{centering}

\protect\caption{\label{fig:2}Energy efficiency versus the reliability thresholds
for communication, $\gamma_{c}$, and radar, $\gamma_{r}$, when $\gamma_{c}=\gamma_{r}=\gamma\textrm{ \ensuremath{\left(dB\right)}}$
for different values of the target height.}

\end{figure*}

Fig. \ref{fig:2} illustrates the energy efficiency versus the reliability
thresholds of communication system, $\gamma_{c}$, and radar system,
$\gamma_{r}$, when $\gamma_{c}=\gamma_{r}=\gamma$ for different
values of the target altitude. Fig \ref{fig:2a} presents the energy
efficiency when the target altitude $h_{t}=1.5\textrm{ m}$ and Fig.
\ref{fig:2b} shows the energy efficiency when the target altitude
$h_{t}=50\textrm{ m}$. The results reveal that, the energy efficiency
degrades with increasing the reliability thresholds, and there exist
an optimal value of the reliability thresholds that maximizes the
energy efficiency.  Clearly, the optimal values for the ISAC network
are distinct to the communications-only network.

\begin{figure*}
\noindent \begin{centering}
\subfloat[\label{fig:3a}Energy efficiency versus the reliability threshold
for communication, $\gamma_{c}$, \\
when $\gamma_{r}=2\textrm{ dB}$
and $h_{t}=1.5m$.]{\noindent \begin{centering}
\includegraphics[scale=0.6]{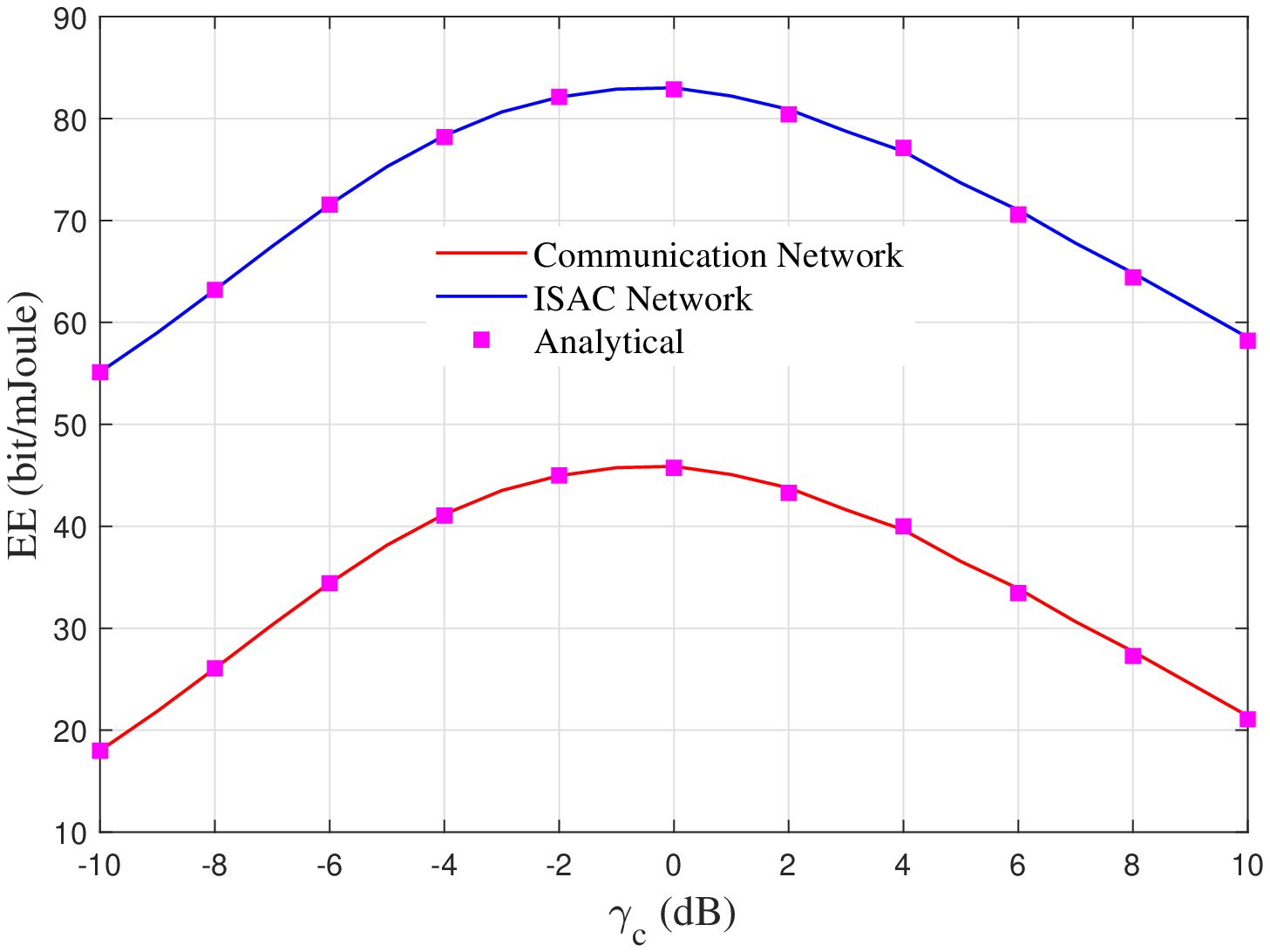}
\par\end{centering}

}\subfloat[\label{fig:3b}Energy efficiency versus the reliability threshold
for communication, $\gamma_{c}$, when $\gamma_{r}=2\textrm{ dB}$
and $h_{t}=50m$.]{\noindent \begin{centering}
\includegraphics[scale=0.6]{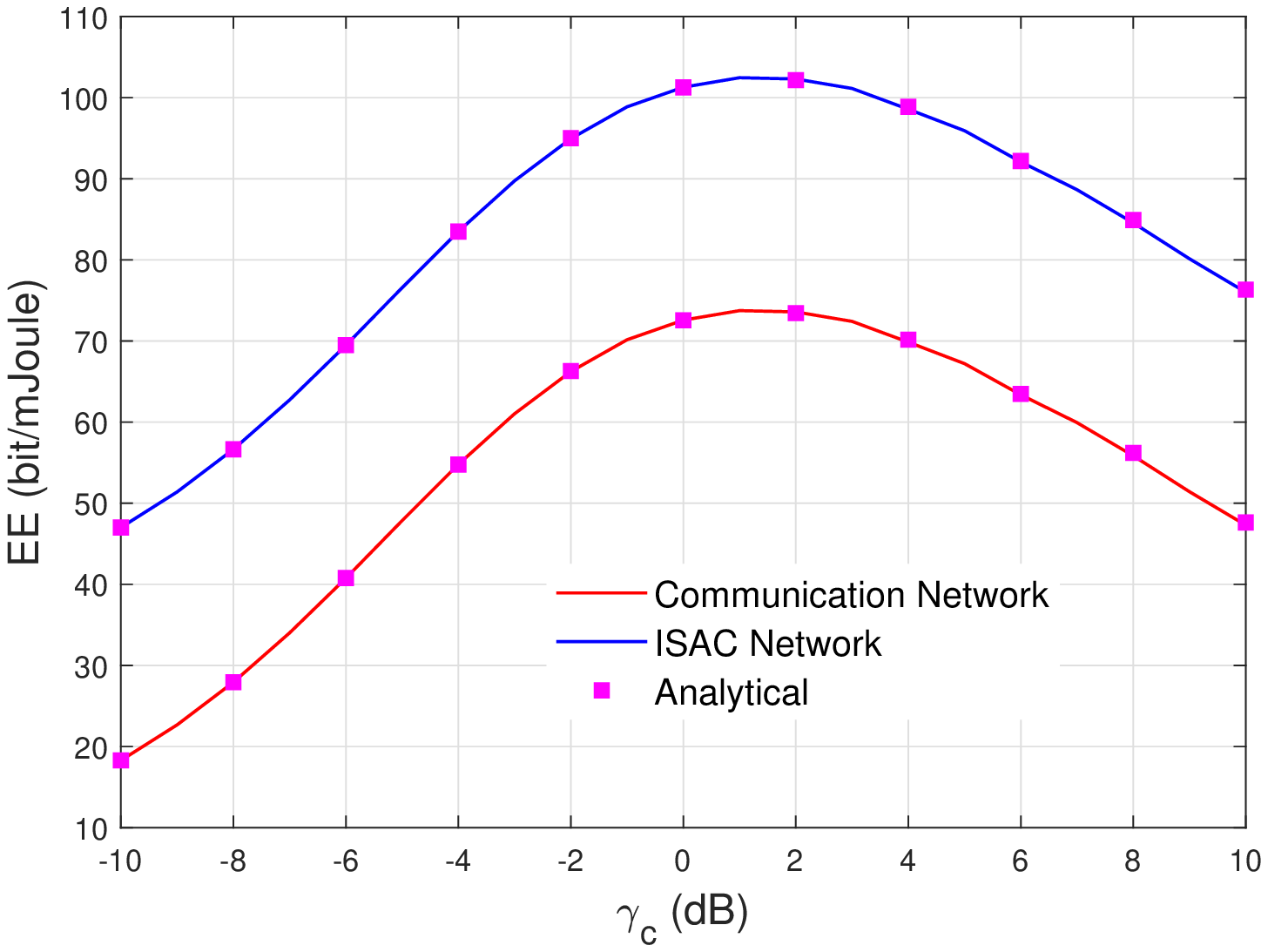}
\par\end{centering}

}
\par\end{centering}

\noindent \begin{centering}
\subfloat[\label{fig:3c}Energy efficiency versus the reliability threshold
for radar, $\gamma_{r}$, \\
when $\gamma_{c}=2\textrm{ dB}$ and $h_{t}=1.5m$.]{\noindent \begin{centering}
\includegraphics[scale=0.55]{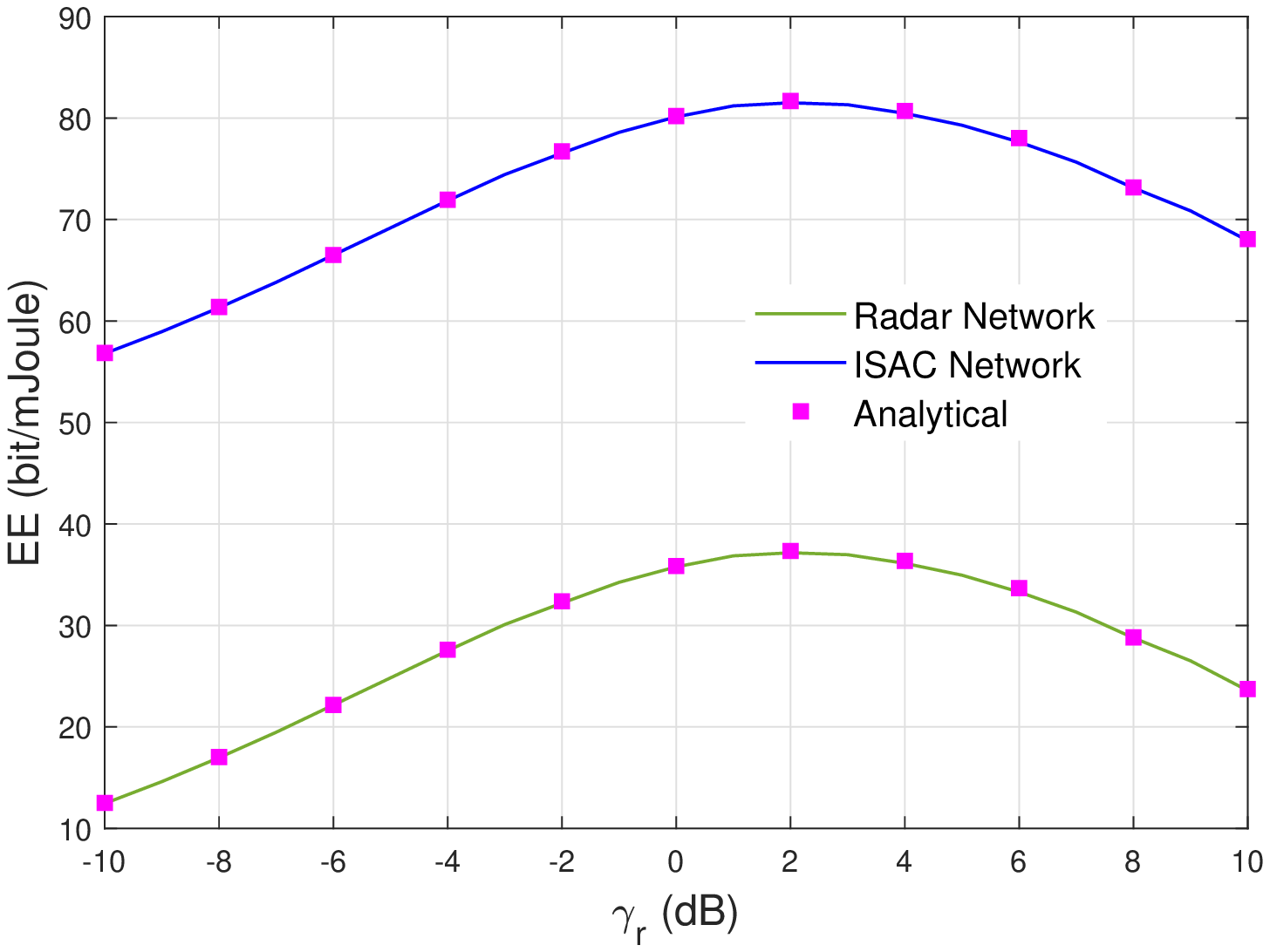}
\par\end{centering}

}\subfloat[\label{fig:3d}Energy efficiency versus the reliability threshold
for radar, $\gamma_{r}$, when $\gamma_{c}=2\textrm{ dB}$ and $h_{t}=50m$.]{\noindent \begin{centering}
\includegraphics[scale=0.55]{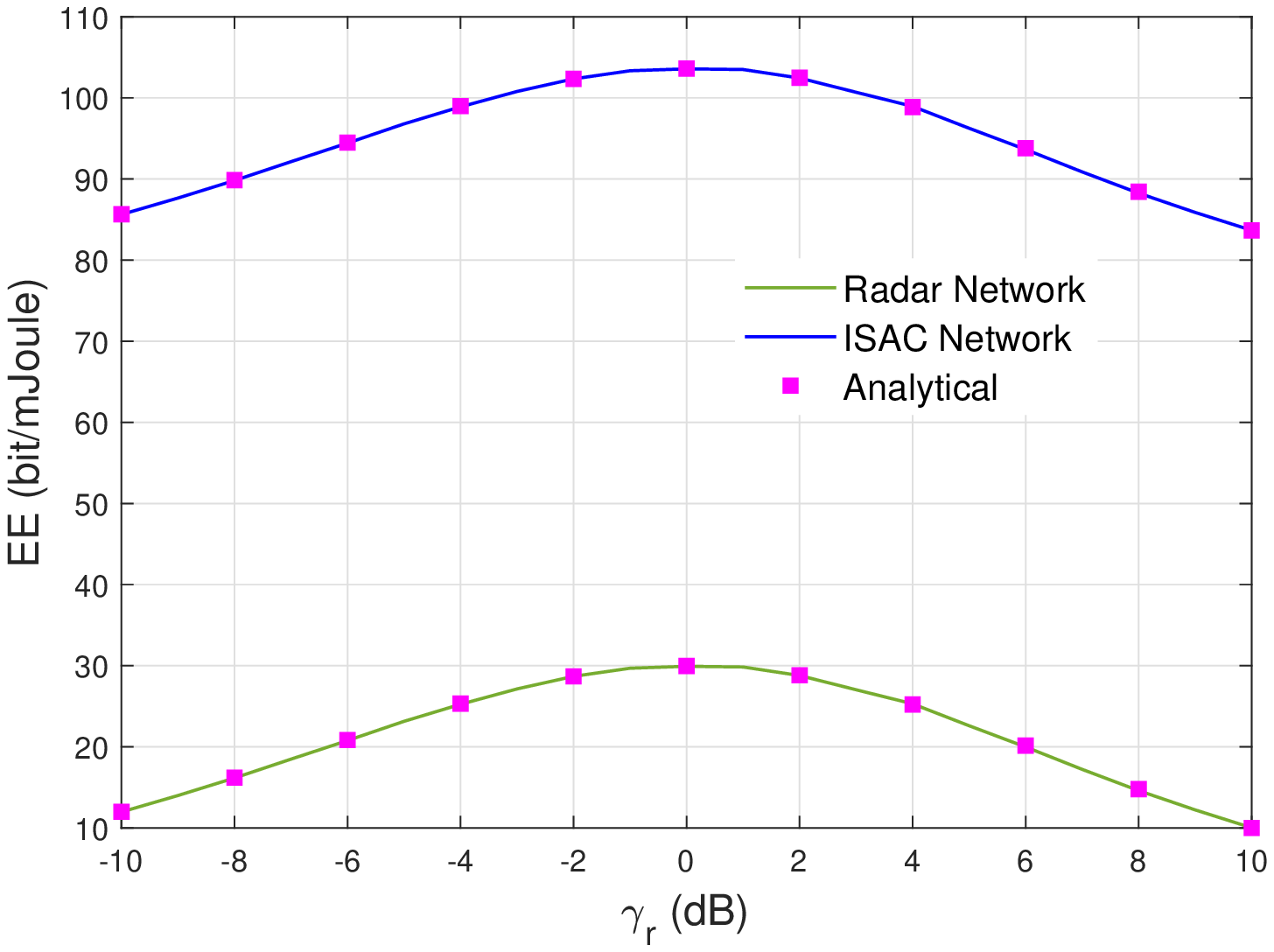}
\par\end{centering}

}
\par\end{centering}

\protect\caption{\label{fig:3}Energy efficiency versus the reliability thresholds
for communication, $\gamma_{c}$, and radar, $\gamma_{r}$, when $\gamma_{c}\protect\neq\gamma_{r}$
for different values of the target height.}
\end{figure*}

In order to investigate more the impact of the reliability thresholds
of communication system, $\gamma_{c}$, and radar system, $\gamma_{r}$,
on the system performance, the energy efficiency is plotted in Fig.
\ref{fig:3} with respect to $\gamma_{c}$ and $\gamma_{r}$, when
$\gamma_{c}\neq\gamma_{r}$ for different target heights. The energy
efficiency versus $\gamma_{c}$ when the target altitude $h_{t}=1.5m$,
$\gamma_{r}=2\textrm{ dB}$  is presented in Fig \ref{fig:3a} and
when the target altitude $h_{t}=50m$, $\gamma_{r}=2\textrm{ dB}$
 is shown in Fig. \ref{fig:3b}. In addition, we plot the energy
efficiency versus $\gamma_{r}$ when the target altitude $h_{t}=1.5m$,
$\gamma_{c}=2\textrm{ dB}$  in Fig \ref{fig:3c} and when the target
altitude $h_{t}=50m$, $\gamma_{c}=2\textrm{ dB}$ in Fig. \ref{fig:3d}.
As we can observe from the figures that the optimal values of the
reliability thresholds for communication/radar systems and DFRC system
are similar, and depend on the target altitude. 

\begin{figure}
\noindent \begin{centering}
\includegraphics[scale=0.55]{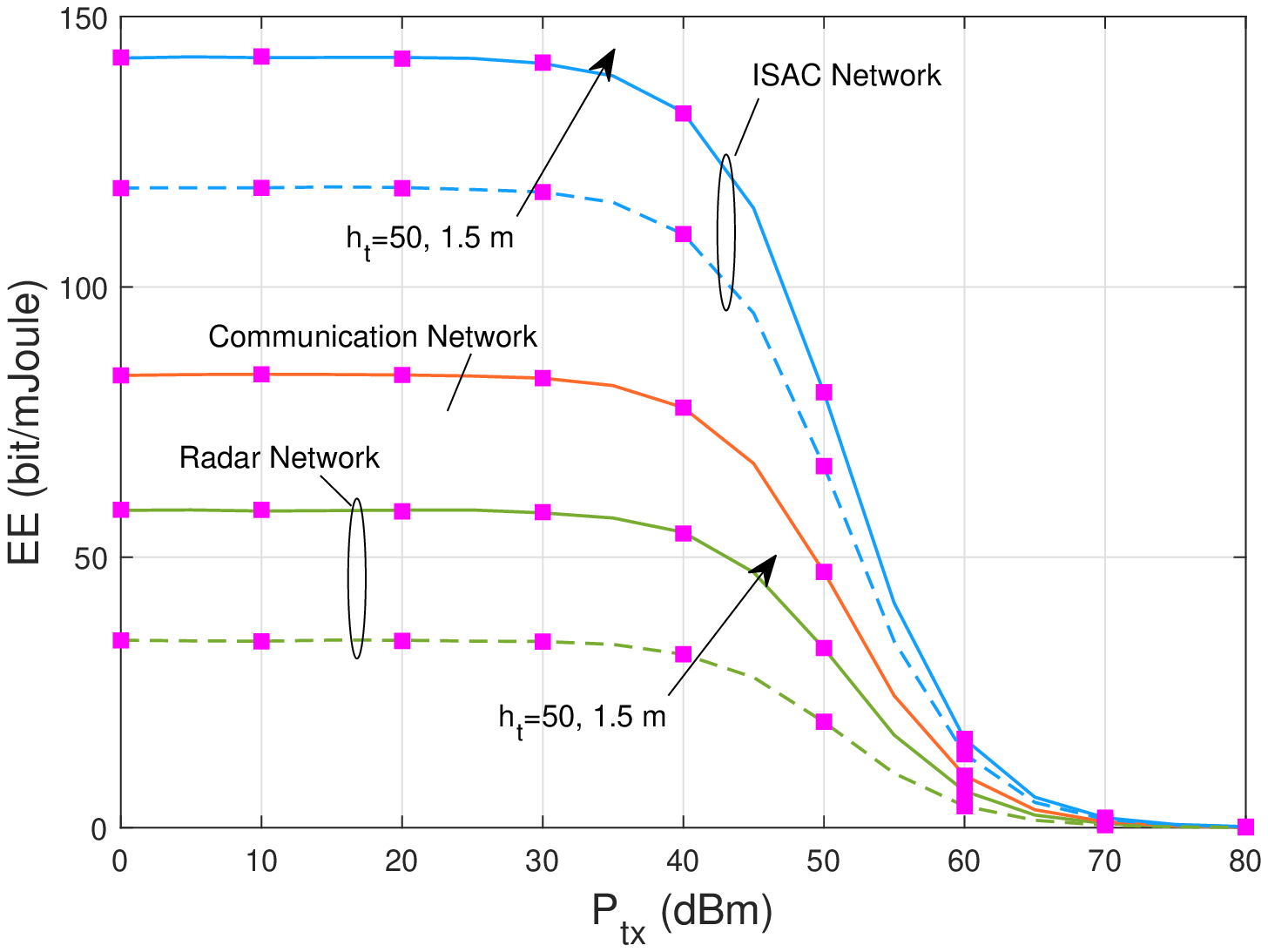}
\par\end{centering}

\protect\caption{\label{fig:4}Energy efficiency versus the transmit power for different
values of the target height.}

\end{figure}

To clearly demonstrate the impact of the transmit power on the system
performance, we plot in Fig. \ref{fig:4} the energy efficiency versus
$P_{tx}$ for different values of $h_{t}$. The good matching between
the analytical and simulation results again confirms the accuracy
of the analysis in this paper. It is evident from these results that
the energy efficiency is flat in low and medium transmit power values
up to $35$dBm. However, the energy efficiency degrades sharply in
high transmit power values when $P_{tx}>35$dBm and goes to zero in
very high $P_{tx}$ values $P_{tx}>60$ dBm. Accordingly, all $P_{tx}$
values up to 35 dBm are optimal and there is no a unique optimal $P_{tx}$
value.\textcolor{blue}{{} This can be justified by the fact that in
the considered scenario the interference power is much higher than
the noise power, and thus the noise power can be neglected. Indeed
this fact leads to eliminate the impact of $P_{tx}$ on the SINR expression based on the EE definition in (1).}
Last observation and as anticipated that, the energy efficiency degrades
as the height of the target increases. 

\begin{figure}
\noindent \begin{centering}
\includegraphics[scale=0.6]{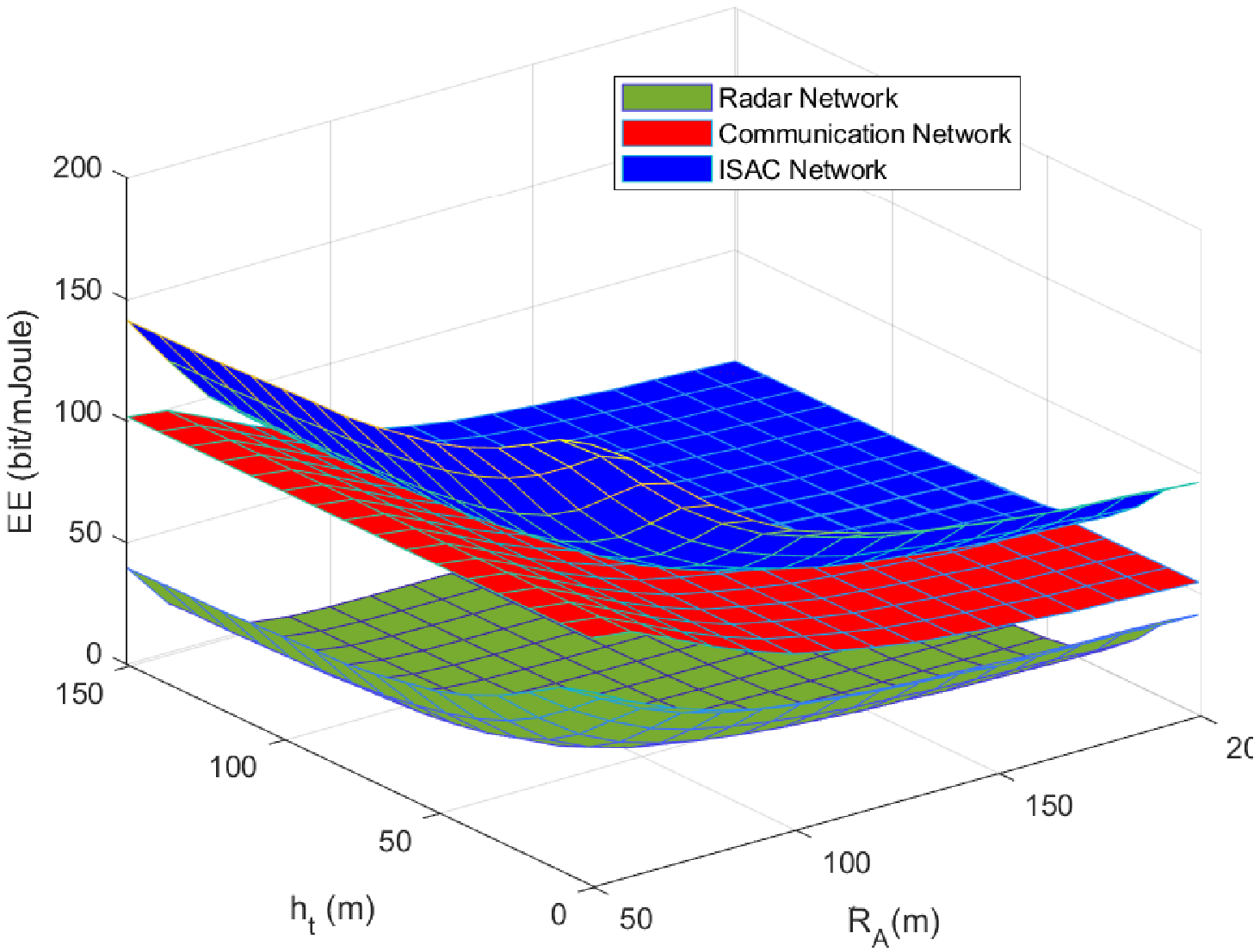}
\par\end{centering}

\protect\caption{\label{fig:5}Energy efficiency of radar and DFRC systems versus the
area radius and the target height.}

\end{figure}

Finally, Fig. \ref{fig:5} depicts a 3D surface plot for the energy
efficiency as a function of the target height $h_{t}$ and the radius
of the area $R_{A}$ for radar system and DFRC system. The common
observation one can see in the two systems is that the energy efficiency
is at its minimum when $R_{A}$ and $h_{t}$ are large. This is because
increasing the area $R_{A}$ results in increasing number of BSs which
leads to increase the interference power. In addition, when the target
is far away from the BS, the received signal will be too weak due
to larger path loss.

\section{Conclusion\label{sec:Conclusion}}

\textcolor{blue}{In this paper we have presented a first network level
study of the design of an ISAC network, using stochastic geometry.
We derived new closed form analytical expressions of the potential
spectral efficiency of communication and radar networks. Then the
new derived expressions of the PSE have
been applied to the analysis and optimization of the energy efficiency
of DFRC networks. Numerical results were presented to confirm our
analysis and to prove the usefulness of the proposed framework for
optimizing the network planning and deployment of DFRC networks. It
has been proved that, the network design changes with the inclusion
of the sensing functionality. The optimal DFRC BS density for the
ISAC network is different to the classical communications-only network,
and this optimal value becomes smaller when the target altitude increases.
Therefore, few number of BSs is enough to achieve the optimal performance
in very high altitude cases. The study in this work can be extended
to design ISAC networks with optimal signalling and precoding schemes.
In addition, the analysis can be straight forward applied to the cases
when the communication users are in uplink mode. The more general cases considering the partially loaded BSs are worthwhile future works.}

\section*{Appendix A}

This appendix derives the PSE of communication system. Firstly, the
coverage probability can be calculated by 
\begin{equation}
\textrm{Pr}\left(\gamma_{o}>\gamma_{c}\right)=\textrm{Pr}\left(\frac{Pd_{o,o}^{-\alpha}\varsigma}{I_{r}+\sigma_{o}^{2}}>\gamma_{c}\right)\label{eq:9},
\end{equation}
\noindent where $I_{r}=P\underset{l\in\Phi_{b}\setminus b}{\sum}d_{l,o}^{-m}\left\Vert \mathbf{h}_{l,o}^{\dagger}\mathbf{W}_{l}\right\Vert ^{2}$.
The expression in (\ref{eq:9}) can be written as  
\begin{gather}
\textrm{Pr}\left(\gamma_{o}>\gamma_{c}\right)=\textrm{Pr}\left(I_{r}<\frac{Pd_{o,o}^{-\alpha}}{\gamma_{c}}\varsigma-\sigma_{o}^{2}\right),
\end{gather}
which can be expressed as,
\begin{equation}
\textrm{Pr}\left(\gamma_{o}>\gamma_{c}\right)=\textrm{Pr}\left(\underset{l\in\Phi_{b}\setminus b}{\sum}d_{l,o}^{-\alpha}\left\Vert \mathbf{h}_{l,o}^{\dagger}\mathbf{W}_{l}\right\Vert ^{2}<\frac{d_{o,o}^{-\alpha}}{\gamma_{c}}\varsigma-\frac{\sigma_{o}^{2}}{P}\right).
\end{equation}
The cumulative distribution function (CDF) of $I_{r}$,
can be obtained with the inverse Laplace transform. The MGF of the
aggregate interference, $I_{r}$, can be calculated by
\begin{equation}
\mathcal{M}_{I_{r}}\left(s\right)=\mathbb{E}\left\{ e^{-s\underset{l\in\Phi_{b}\setminus b}{\sum}d_{l,o}^{-\alpha}\left\Vert \mathbf{h}_{l,o}^{\dagger}\mathbf{W}_{l}\right\Vert ^{2}}\right\} .
\end{equation}

Then, the MGF can be found as
\begin{equation}
\mathcal{M}_{I_{r}}\left(s\right)=\mathbb{E}_{\Phi_{b}}\left\{ \underset{l\in\Phi_{b}\setminus b}{\prod}\mathbb{E}_{Y_{l}}\left\{ e^{-sd_{l,o}^{-\alpha}\left\Vert \mathbf{h}_{l,o}^{\dagger}\mathbf{W}_{l}\right\Vert ^{2}}\right\} \right\} .
\end{equation}

Now, $Y_{l}=\left\Vert \mathbf{h}_{l,o}^{\dagger}\mathbf{W}_{l}\right\Vert ^{2}$
has gamma distribution, i.e., $Y_{l}\sim\Gamma\left(\beta,\alpha\right)$,
with $\beta=K,\alpha=1$. Therefore, 
\begin{equation}
\mathcal{M}_{I_{r}}\left(s\right)=\mathbb{E}_{\Phi_{b}}\left\{ \underset{l\in\Phi_{b}\setminus b}{\Pi}\left(1-\frac{sd_{l,o}^{-\alpha}}{\beta}\right)^{-\alpha}\right\} .
\end{equation}

Using the probability generating functional (PGFL) of PPP \cite{book},
$\mathcal{M}_{I_{r}}\left(s\right)$ can be expressed as
\begin{equation}
\mathcal{M}_{I_{r}}\left(s\right)=\exp\left(-\lambda_{b}\pi\stackrel[d]{r}{\int}\left(1-\left(1-\frac{s\gamma_{c}u^{-\frac{\alpha}{2}}}{\beta}\right)^{-\tilde{\alpha}}\right)du\right),
\end{equation}
which can be found as
\[
\mathcal{M}_{I_{r}}\left(s\right)\!=\!\exp\left(-\lambda_{b}\!\left(-d+r+d\ensuremath{_{2}F_{1}}\left[\alpha,-\frac{2}{\alpha},\frac{-2+\alpha}{\alpha},\frac{sd^{-\alpha/2}}{\beta}\right]\right.\right.
\]
\begin{equation}
\left.\left.-r\ensuremath{_{2}F_{1}}\left[\alpha,-\frac{2}{\alpha},\frac{-2+\alpha}{\alpha},\frac{sr^{-\alpha/2}}{\beta}\right]\right)\right)\label{eq:16-1}.
\end{equation}

If the MGF is invertible, the CDF can be found as \cite{mgfcdf,menoma}
\begin{equation}
F\left(x\right)=\frac{1}{2\pi j}\stackrel[a-j\infty]{a+j\infty}{\int}\mathcal{M}_{x}\left(s\right)e^{sx}ds,
\end{equation}
where $\mathcal{M}_{x}\left(s\right)$ is the MGF of $x$. which has
been simplified in \cite{mgfcdf,menoma} to
\begin{align}
	F_{X}\left(x\right)=&\dfrac{2^{-Q}e^{\frac{A}{2}}}{x}\stackrel[q=0]{Q}{\sum}\left(\begin{array}{c}
		Q\\
		q
	\end{array}\right)\stackrel[n=0]{N+q}{\sum}\dfrac{\left(-1\right)^{n}}{\beta_{n}} \nonumber \\
	&\times\mathfrak{R}\left(\dfrac{\mathcal{M}_{X}\left(\dfrac{A+2\pi jn}{2x}\right)}{\dfrac{A+2\pi jn}{2x}}\right)+E(A,N,Q)\label{eq:91-1},
\end{align}
where $j^{2}=-1$, $\mathfrak{R}\left\{ .\right\} $ denotes the real
part; $A$, $N$ and $Q$ are positive integers used to control accuracy
and satisfy the condition that a remainder error term $E(A,N,Q)$
is negligible compared with the first term,
\[
\left|E(A,N,Q)\right|=\dfrac{e^{-A}}{1-e^{-A}}+\left|\dfrac{2^{-Q}e^{\frac{A}{2}}}{\gamma_{M}}\stackrel[q=0]{Q}{\sum}\left(-1\right)^{N+1+q}\right.
\]
\begin{align}
\left.\left(\begin{array}{c}
Q\\
q
\end{array}\right)\mathfrak{R}\left(\dfrac{\mathcal{M}_{\varphi}\left(\dfrac{A+2\pi j\left(N+q+1\right)}{2\gamma_{M}}\right)}{\dfrac{A+2\pi j\left(N+q+1\right)}{2\gamma_{M}}}\right)\right|,\label{eq:91-2-1}
\end{align}
\[
\beta_{n}=\begin{cases}
\begin{array}{l}
2\qquad\qquad n=0\\
1\qquad n=1,2,\cdots,N+q
\end{array},\end{cases}
\]
By plug in (\ref{eq:16-1}) into (\ref{eq:91-1}) we can get
\[
\textrm{Pr}\left(\underset{l\in\Phi_{b}\setminus b}{\sum}d_{l,o}^{-m}\left\Vert \mathbf{h}_{l,o}^{\dagger}\mathbf{W}_{l}\right\Vert ^{2}<\frac{d_{o,o}^{-\alpha}}{\gamma_{c}}\right)=\dfrac{2^{-Q}e^{\frac{A}{2}}}{\frac{d_{o,o}^{-\alpha}}{\gamma_{c}}}\stackrel[q=0]{Q}{\sum}\left(\begin{array}{c}
Q\\
q
\end{array}\right)
\]
\begin{equation}
\stackrel[n=0]{N+q}{\sum}\dfrac{\left(-1\right)^{n}}{\beta_{n}}\mathfrak{R}\left(\dfrac{\mathcal{M}_{I_{r}}\left(\dfrac{A+2\pi jn}{2\frac{d_{o,o}^{-\alpha}}{\gamma_{c}}}\right)}{\dfrac{A+2\pi jn}{2\frac{d_{o,o}^{-\alpha}}{\gamma_{c}}}}\right)+E(A,N,Q).\label{eq:91}.
\end{equation}
Additionally, it is worthy mentioning that, the CDF of the aggregate
interference in terms of its characteristic function (CF) can be also
obtained by invoking the Gil\textendash Pelaez inversion theorem in
\cite{Pelaez}. In this case the CDF can derived by
\begin{equation}
F\left(x\right)=\frac{1}{2}+\frac{1}{2\pi}\stackrel[0]{\infty}{\int}\frac{e^{itx}\phi\left(-t\right)-e^{-itx}\phi\left(t\right)}{it}dt\label{eq:27},
\end{equation}
where $\phi\left(t\right)$ is CF. However, all the aforementioned
ways to derive the coverage probability are complicated and simple
analytical expression of the coverage probability is hard to obtain.

\textcolor{blue}{In order to obtain simple closed-form expression,
very tight approximation has been considered in the literature for
such complicated scenarios. It is also a very popular technique because
of its simplicity and accuracy. This is based on deriving the coverage
probability by only considering the subset of dominant interferences.}
 Hence we have
\begin{equation}
\textrm{Pr}\left(\gamma_{o}>\gamma_{c}\right)=\textrm{Pr}\left(I_{r}<\frac{Pd_{o,o}^{-\alpha}}{\gamma_{c}}\varsigma-\sigma_{o}^{2}\right)\label{eq:52},
\end{equation}
and 
\begin{equation}
\textrm{Pr}\left(I_{r}>\frac{Pd_{o,o}^{-\alpha}}{\gamma_{c}}\varsigma-\sigma_{o}^{2}\right)=1-\textrm{Pr}\left(I_{r}<\frac{Pd_{o,o}^{-\alpha}}{\gamma_{c}}\varsigma-\sigma_{o}^{2}\right).
\end{equation}

The probability term in the right-hand side (RHS) is equal to

\begin{equation}
\textrm{Pr}\left(I_{r}<\frac{Pd_{o,o}^{-\alpha}}{\gamma_{c}}\varsigma-\sigma_{o}^{2}\right)=\textrm{Pr}\left(\Phi_{b}={\phi}\right).
\end{equation}

Thus to compute the probability we can observe that the event $\textrm{Pr}\left(I_{r}<\frac{Pd_{o,o}^{-m}}{\gamma_{c}}\varsigma-\sigma_{o}^{2}\right)$
is the same as the event $\textrm{Pr}\left(\Phi_{b}={\phi}\right)$.
With this observation using the expression for the void probability
of a Poisson process, the conditional probability can be written as 

\begin{align}
	&\textrm{Pr}\left(I_{r}<\frac{Pd_{o,o}^{-\alpha}}{\gamma_{c}}\varsigma-\sigma_{o}^{2}\right) \nonumber \\
	 =&e^{\left(-\stackrel[d_{o,o}]{\infty}{\int}\textrm{Pr}\left(\left|x_{l}\right|^{-\alpha}\left\Vert \mathbf{h}_{l,o}^{\dagger}\mathbf{W}_{l}\right\Vert ^{2}>\frac{d_{o,o}^{-\alpha}}{\gamma_{c}}\varsigma-\frac{\sigma_{0}^{2}}{P}\right)\right)} \nonumber \\
	=&e^{\left(-\stackrel[d_{o,o}]{\infty}{\int}\textrm{Pr}\left(\left\Vert \mathbf{h}_{l,o}^{\dagger}\mathbf{W}_{l}\right\Vert ^{2}>\left(\frac{d_{o,o}^{-\alpha}}{\gamma_{c}}\varsigma-\frac{\sigma_{0}^{2}}{P}\right)\left|x_{l}\right|^{\alpha}\right)dx\right)}  \nonumber \\
	=&e^{\left(-2\pi\lambda\stackrel[d_{o,o}]{\infty}{\int}\textrm{Pr}\left(\left\Vert \mathbf{h}_{l,o}^{\dagger}\mathbf{W}_{l}\right\Vert ^{2}>tr^{\alpha}\right)rdr\right)}\label{eq:55},
\end{align}

\noindent where $t=\left(\frac{d_{o,o}^{-m}}{\gamma_{c}}\varsigma-\frac{\sigma_{0}^{2}}{P}\right).$
Since $Y_{l}=\left\Vert \mathbf{h}_{l,o}^{\dagger}\mathbf{W}_{l}\right\Vert ^{2}$
has gamma distribution$Y_{l}\overset{d}{\backsim}\Gamma\left(\kappa,\beta\right)$,
(\ref{eq:55}) can be expressed as 
\begin{align}
	&\textrm{Pr}\left(I_{r}<\frac{Pd_{o,o}^{-m}}{\gamma_{c}}\varsigma-\sigma_{0}^{2}\right) \nonumber\\
	=&e^{\left(-2\pi\lambda\stackrel[d_{o,o}]{\infty}{\int}\left(1-\left(\frac{\gamma\left(\kappa,\beta tr^{\alpha}\right)}{\Gamma\left(\kappa,\right)}\right)\right)rdr\right)} \nonumber  \\
	=& e^{\left(-2\pi\lambda\stackrel[d_{o,o}]{\infty}{\int}\left(1-\left(1-\frac{\Gamma\left(\kappa,\beta tr^{\alpha}\right)}{\Gamma\left(\kappa\right)}\right)\right)rdr\right)}\nonumber  \\
	=&e^{\left(-2\pi\lambda\stackrel[d_{o,o}]{\infty}{\int}\left(\frac{\Gamma\left(\kappa,\beta tr^{\alpha}\right)}{\Gamma\left(\kappa\right)}\right)rdr\right)}\label{eq:65}.
\end{align}
Using the fact that, $\Gamma\left(x,y\right)=\left(x-1\right)!e^{-y}\stackrel[k=0]{x-1}{\sum}\frac{y^{k}}{k!}$,
(\ref{eq:65}) can be rewritten as (\ref{EquationLong4}). Now by taking the average over $d_{o,o}$ we can obtain equation (\ref{eq:58}), as shown at the top of the page. Since the pdf of $d_{o,o}$ is given by $f_{do,o}=2\pi\lambda_{b}re^{-\pi\lambda r^{2}}$,
the average in (\ref{eq:58}) can be found as (\ref{EquationLong5}).
\begin{figure*}
	\begin{equation}\label{EquationLong4}
		\begin{aligned}
			&\textrm{Pr}\left(I_{r}<\frac{Pd_{o,o}^{-m}}{\gamma_{c}}\varsigma-\sigma_{0}^{2}\right) =e^{\left(-\frac{2\pi\lambda\left(\kappa-1\right)!}{\Gamma\left(\kappa\right)}\stackrel[k=0]{\kappa-1}{\sum}\frac{1}{k!}\stackrel[d_{o,o}]{\infty}{\int}\left(e^{-\beta tr^{\alpha}}\left(\beta tr^{\alpha}\right)^{k}\right)rdr\right)} \nonumber \\
			&= e^{-\frac{2\pi\lambda\left(\kappa-1\right)!}{\Gamma\left(\kappa\right)}\stackrel[k=0]{\kappa-1}{\sum}\frac{1}{\alpha\, k!}\left(\left(\beta\left(\left(\frac{d_{o,o}^{-\alpha}}{\gamma_{c}}-\frac{\sigma_{0}^{2}}{P}\right)\right)\right)^{\frac{-2}{\alpha}}\right) \times \Gamma\left(\frac{k+2}{m},\beta\left(\left(\frac{d_{o,o}^{-\alpha}}{\gamma_{c}}-\frac{\sigma_{0}^{2}}{P}\right)\right)\right)}.
		\end{aligned}
	\end{equation}
\end{figure*}

\begin{figure*}
	\begin{equation}
		\textrm{Pr}\left(I_{r}<\frac{Pd_{o,o}^{-\alpha}}{\gamma_{c}}\varsigma-\sigma_{0}^{2}\right)=E_{d_{o,o}}\left(e^{\left(-\frac{2\pi\lambda\left(\kappa-1\right)!}{\Gamma\left(\kappa\right)}\stackrel[k=0]{\kappa-1}{\sum}\frac{1}{\alpha\, k!}\left(\left(\beta\left(\left(\frac{d_{o,o}^{-\alpha}}{\gamma_{c}}-\frac{\sigma_{0}^{2}}{P}\right)\right)\right)^{\frac{-2}{\alpha}}\right)\Gamma\left(\frac{k+2}{\alpha},\beta\left(\left(\frac{d_{o,o}^{-\alpha}}{\gamma_{c}}-\frac{\sigma_{0}^{2}}{P}\right)\right)\right)\right)}\right)\label{eq:58}.
	\end{equation}
\end{figure*}

\begin{figure*}
\begin{equation}\label{EquationLong5}
	\begin{aligned}
		\textrm{Pr}\left(I_{r}<\frac{Pd_{o,o}^{-\alpha}}{\gamma_{c}}\varsigma-\sigma_{0}^{2}\right)&=2\pi\lambda\stackrel[0]{\infty}{\int}re^{-\pi\lambda r^{2}}e^{\left(-\frac{2\pi\lambda\left(\kappa-1\right)!}{\Gamma\left(\kappa\right)}\stackrel[k=0]{\kappa-1}{\sum}\frac{1}{\alpha\, k!}\left(\beta\left(\frac{r^{-\alpha}}{\gamma_{c}}-\frac{\sigma_{0}^{2}}{P}\right)\right)^{\frac{-2}{\alpha}}\Gamma\left(\frac{k+2}{\alpha},\beta\left(\frac{r^{-\alpha}}{\gamma_{c}}-\frac{\sigma_{0}^{2}}{P}\right)\right)\right)}dr \\
		\hphantom{\textrm{Pr}\left(I_{r}<\frac{Pd_{o,o}^{-m}}{\gamma_{c}}\varsigma-\sigma_{0}^{2}\right)}&=2\pi\lambda\stackrel[0]{\infty}{\int}re^{-\pi\lambda r^{2}}\stackrel[k=0]{\kappa-1}{\Pi}e^{-\frac{2\pi\lambda\left(\kappa-1\right)!}{\Gamma\left(\kappa\right)}\left(\frac{1}{\alpha\, k!}\left(\left(\beta\left(\frac{r^{-\alpha}}{\gamma_{c}}-\frac{\sigma_{0}^{2}}{P}\right)\right)^{\frac{-2}{\alpha}}\right)\Gamma\left(\frac{k+2}{\alpha},\beta\left(\frac{r^{-\alpha}}{\gamma_{c}}-\frac{\sigma_{0}^{2}}{P}\right)\right)\right)}dr \\
		\hphantom{\textrm{Pr}\left(I_{r}<\frac{Pd_{o,o}^{-m}}{\gamma_{c}}\varsigma-\sigma_{0}^{2}\right)}&=2\pi\lambda\stackrel[0]{\infty}{\int}r\stackrel[k=0]{\kappa-1}{\Pi}e^{-\left(\frac{2\pi\lambda\left(\kappa-1\right)!}{\Gamma\left(\kappa\right)}\left(\frac{\left(\beta\left(\frac{r^{-\alpha}}{\gamma_{c}}-\frac{\sigma_{0}^{2}}{P}\right)\right)^{\frac{-2}{\alpha}}}{\alpha\, k!}\Gamma\left(\frac{k+2}{\alpha},\beta\left(\frac{r^{-\alpha}}{\gamma_{c}}-\frac{\sigma_{0}^{2}}{P}\right)\right)\right)+\frac{\pi\lambda r^{2}}{\kappa}\right)}dr
	\end{aligned}.
\end{equation}
\end{figure*}

This expression can be simplified using Gaussian Quadrature rules.
Thus, the coverage probability can be written as (\ref{eq:16}). 
\begin{figure*}
\begin{equation}
	\begin{aligned}
		&\textrm{Pr}\left(I_{r}<\frac{Pd_{o,o}^{-\alpha}}{\gamma_{c}}\varsigma-\sigma_{0}^{2}\right)=2\pi\lambda\stackrel[i=1]{n}{\sum}\textrm{H}_{i}e^{r_{i}}r_{i} \\
		& \stackrel[k=0]{\kappa-1}{\Pi}e^{-\left(\frac{2\pi\lambda\left(\kappa-1\right)!}{\Gamma\left(\kappa\right)}\left(\frac{\left(\left(\beta\left(\frac{r_{i}^{-\alpha}}{\gamma_{c}}-\frac{\sigma_{0}^{2}}{P}\right)\right)^{\frac{-2}{\alpha}}\right)}{\alpha\, k!}\Gamma\left(\frac{k+2}{\alpha},\beta\left(\frac{r_{i}^{-\alpha}}{\gamma_{c}}-\frac{\sigma_{0}^{2}}{P}\right)\right)\right)+\frac{\pi\lambda r_{i}^{2}}{\kappa}\right)}+R_{i},\label{eq:16}
	\end{aligned}
\end{equation}
\end{figure*}
\noindent where $r_{i}$ and $\textrm{H}_{i}$ are the $i^{th}$ zero
and the weighting factor of the Laguerre polynomials, respectively,
and the remainder $R_{i}$ is negligible for $n>15$ \cite{book2}.
 Finally, by substituting (\ref{eq:16})/(\ref{eq:52}) into (\ref{eq:3})
we can find the PSE of communication system presented in Theorem 1.

\section*{Appendix B}

This appendix derives the PSE of radar system. Now the coverage probability
of the radar system can be written as 

\[
\textrm{Pr}\left(\gamma_{b}>\gamma_{r}\right)=\textrm{Pr}\left(\frac{\frac{\left|\alpha_{b}\right|^{2}P\mathbf{w}_{r}^{H}\mathbf{G}_{b}\mathbf{W}_{b}\mathbf{s}_{b}\mathbf{s}_{b}^{H}\mathbf{W}_{b}^{H}\mathbf{G}_{b}^{H}\mathbf{w}_{r}}{\mathbf{w}_{r}^{H}\mathbf{w}_{r}}}{\underset{l\in\Phi_{b}\setminus b}{\sum}\frac{P\mathbf{w}_{r}^{H}\mathbf{H}_{l,b}\mathbf{W}_{l}\mathbf{s}_{l}\mathbf{s}_{l}^{H}\mathbf{W}_{l}^{H}\mathbf{H}_{l,b}^{H}\mathbf{w}_{r}}{\mathbf{w}_{r}^{H}\mathbf{w}_{r}}+\sigma_{b}^{2}}>\gamma_{r}\right)
\vspace{-2mm}
\]

\begin{equation}
\hphantom{\textrm{Pr}\left(\gamma_{b}>\gamma_{r}\right)}=\textrm{Pr}\left(X>\left|\alpha_{b}\right|^{-2}\gamma_{r}\underset{l\in\Phi_{b}\setminus b}{\sum}Y_{l}+\left|\alpha_{b}\right|^{-2}P^{-1}\gamma_{r}\sigma_{b}^{2}\right),
\end{equation}

\noindent where $X=\frac{\mathbf{w}_{r}^{H}\mathbf{G}_{b}\mathbf{W}_{b}\mathbf{s}_{b}\mathbf{s}_{b}^{H}\mathbf{W}_{b}^{H}\mathbf{G}_{b}^{H}\mathbf{w}_{r}}{\mathbf{w}_{r}^{H}\mathbf{w}_{r}},\,\textrm{and }Y_{l}=\frac{\mathbf{w}_{r}^{H}\mathbf{H}_{l,b}\mathbf{W}_{l}\mathbf{s}_{l}\mathbf{s}_{l}^{H}\mathbf{W}_{l}^{H}\mathbf{H}_{l,b}^{H}\mathbf{w}_{r}}{\mathbf{w}_{r}^{H}\mathbf{w}_{r}}$.
In such networks $\underset{l\in\Phi_{b}\setminus b}{\sum}Y_{l}\gg\sigma_{b}^{2}$,
thus

\begin{equation}
\textrm{Pr}\left(\gamma_{b}>\gamma_{r}\right)=\textrm{Pr}\left(X>\left|\alpha_{b}\right|^{-2}\gamma_{r}\underset{l\in\Phi_{b}\setminus b}{\sum}Y_{l}\right)=1-\textrm{Pr}\left(X<\left|\alpha_{b}\right|^{-2}\gamma_{r}\underset{l\in\Phi_{b}\setminus b}{\sum}Y_{l}\right).
\end{equation}

\textcolor{blue}{From the properties of wishart and inverse-wishart
distributions, it has been shown that $\mathbf{W}_{b}\mathbf{S}_{b}\mathbf{W}_{b}^{H}\xrightarrow[\approx]{d}\stackrel[j=1]{n}{\sum}\lambda_{j}\mathbf{IW}_{b,j}$,
where $\xrightarrow[\approx]{d}$ denotes the approximated distribution,
$\lambda_{j}$ is the $j^{th}$ non-zero eigenvalues of $\mathbf{S}_{b}$
and $\mathbf{IW}_{b,j}$ are independent non-central inverse Wishart
(}\textbf{\textcolor{blue}{IW}}\textcolor{blue}{), in case $\mathbf{S}_{b}$
has one eigenvalue we can obtain, $\mathbf{W}_{b}\mathbf{S}_{b}\mathbf{W}_{b}^{H}\xrightarrow[\approx]{d}\lambda\mathbf{IW}_{b}$}
\cite{distnew,distnew2,eaton,bookaspects}. Accordingly, for non zero
vector $\mathbf{z}$ we can get $\frac{\mathbf{z}^{H}\mathbf{W}_{b}\mathbf{S}_{b}\mathbf{W}_{b}^{H}\mathbf{z}}{\mathbf{z}\lambda\mathbf{z}}\xrightarrow[\approx]{d}\mathbf{\textrm{inverse}}\textrm{Gamma}$.
Therefore, by conditioning on $\alpha_{b}$, $X=\frac{\mathbf{w}_{r}^{H}\mathbf{G}_{b}\mathbf{W}_{b}\mathbf{s}_{b}\mathbf{s}_{b}^{H}\mathbf{W}_{b}^{H}\mathbf{G}_{b}^{H}\mathbf{w}_{r}}{\left\Vert \mathbf{I}\left(N,K\right)\mathbf{S}\mathbf{I}\left(K,N\right)\right\Vert _{F}\mathbf{w}_{r}^{H}\mathbf{w}_{r}}$,
has  inverse gamma$\left(N_{t},\alpha\right)$ distribution. Similarly
$Y$ has  inverse gamma$\left(N_{t},\beta\right)$ distribution.
Thus,

\begin{equation}
\textrm{Pr}\left(\gamma_{b}>\gamma_{r}\right)=1-\textrm{Pr}\left(X<\left|\alpha_{b}\right|^{-2}\gamma_{r}\underset{l\in\Phi_{b}\setminus b}{\sum}Y_{l}\right)=1-\textrm{Pr}\left(X<\left|\alpha_{b}\right|^{-2}\gamma_{r}y\right),
\end{equation}

\noindent where $y=\underset{l\in\Phi_{b}\setminus b}{\sum}Y_{l}.$
Since, $X$ has inverse gamma$\left(N_{t},\tilde{\alpha}\right)$
distribution, by conditioning on $y$ we can write 

\begin{equation}
	\textrm{Pr}\left(\gamma_{b}>\gamma_{r}\right)=1-\frac{\Gamma\left(N,\frac{\tilde{\alpha}}{\left|\alpha_{b}\right|^{-2}\gamma_{r}y}\right)}{\Gamma\left(N\right)} =1-\frac{\Gamma\left(N\right)-\gamma\left(N,\frac{\tilde{\alpha}}{\left|\alpha_{b}\right|^{-2}\gamma_{r}y}\right)}{\Gamma\left(N\right)}=-\frac{\gamma\left(N,\frac{\tilde{\alpha}}{\left|\alpha_{b}\right|^{-2}\gamma_{r}y}\right)}{\Gamma\left(N\right)}\label{eq:67}.
\end{equation}

Using the series expression of Gamma function, (\ref{eq:67}) can
be expressed as 
\begin{equation}
\textrm{Pr}\left(\gamma_{b}>\gamma_{r}\right)=\left(1-e^{-\frac{\tilde{\alpha}}{\gamma_{r}y\left|\alpha_{b}\right|^{-2}}}\stackrel[i=0]{N-1}{\sum}\frac{\left(\frac{\tilde{\alpha}}{\left|\alpha_{b}\right|^{-2}\gamma_{r}y}\right)^{i}}{i!}\right),
\end{equation}

and thus 

\begin{equation}
\textrm{Pr}\left(\gamma_{b}<\gamma_{r}\right)=e^{-\frac{\tilde{\alpha}}{\gamma_{r}y\left|\alpha_{b}\right|^{-2}}}\stackrel[i=0]{N-1}{\sum}\frac{\left(\frac{\tilde{\alpha}}{\left|\alpha_{b}\right|^{-2}\gamma_{r}y}\right)^{i}}{i!}.
\end{equation}
Now by taking the average over $y$ we can write,

\begin{equation}
\textrm{Pr}\left(\gamma_{b}<\gamma_{r}\right)=\stackrel[i=0]{N-1}{\sum}\frac{\left(\frac{\tilde{\alpha}}{\gamma_{r}\left|\alpha_{b}\right|^{-2}}\right)^{i}}{i!}\mathbf{E}_{\frac{1}{y}}\left[\left(\frac{1}{y}\right)^{i}e^{-\frac{\tilde{\alpha}}{\gamma_{r}y\left|\alpha_{b}\right|^{-2}}}\right]=\stackrel[i=0]{N-1}{\sum}\frac{\left(s\right)^{i}}{i!}\mathbf{E}_{\frac{1}{y}}\left[\left(\frac{1}{y}\right)^{i}e^{-\frac{s}{y}}\right]\label{eq:71},
\end{equation}

\noindent where $s=\frac{\tilde{\alpha}}{\gamma_{r}\left|\alpha_{b}\right|^{-2}}.$
\textcolor{blue}{Following the property of the Laplace transform,
we have, $E_{\upsilon}\left[x^{n}e^{-s\upsilon}\right]=\left(-1\right)^{n}\frac{d^{n}}{ds^{n}}\mathcal{L}_{\upsilon}\left(s\right)$,
assuming $\upsilon=\frac{1}{y}$,}
\textcolor{blue}{
\begin{equation}
\textrm{Pr}\left(\gamma_{b}<\gamma_{r}\right)=\stackrel[i=0]{N-1}{\sum}\frac{\left(-s\right)^{i}}{i!}\frac{d^{i}}{ds^{i}}\mathcal{L}_{\upsilon}\left(s\right)\label{eq:72},
\end{equation}
}
\textcolor{blue}{where $\mathcal{L}_{\upsilon}\left(s\right)=E_{\upsilon}\left[e^{-\frac{s}{\text{y}}}\right]$
and can be found as $\mathcal{L}_{\upsilon}\left(s\right)=E_{\upsilon}\left[e^{-\frac{s}{\text{\ensuremath{\underset{l\in\Phi_{b}\setminus b}{\sum}Y_{l}}}}}\right]$,
using Jensen inequality we can get approximation simple closed form
$\mathcal{L}_{\upsilon}\left(s\right)\triangleq\left[e^{-\frac{s}{E_{\Phi,y}\text{\ensuremath{\left[\underset{l\in\Phi_{b}\setminus b}{\sum}Y_{l}\right]}}}}\right]$.
Following from the i.i.d. distribution of $y_{l}$ and its further
independence from the point process $\Phi$, and applying Campbell's 
theory, the Laplace transform can be obtained }as

\textcolor{blue}{
\begin{equation}
	\begin{aligned}
		\mathcal{L}_{\upsilon}\left(s\right) & \triangleq\left[e^{-\frac{s}{E_{\Phi}\text{\ensuremath{\left[\underset{l\in\Phi_{b}\setminus b}{\sum}E\left[y_{l}\right]\right]}}}}\right] \hphantom{\mathcal{L}_{x}\left(s\right)}\triangleq e^{-\frac{s}{2\pi\lambda_{b}\stackrel[r]{\infty}{\smallint}\left(\stackrel[0]{\infty}{\smallint}y_{l}f_{y_{l}\left(y\right)}dy\right)vdv}}=e^{-\frac{s}{2\pi\lambda_{b}\stackrel[r]{\infty}{\smallint}\frac{\beta\left|v\right|^{-\tilde{\alpha}}}{N-1}vdv}} \\
		 \hphantom{\mathcal{L}_{x}\left(s\right)}&\triangleq e^{-\frac{s}{2\pi\lambda_{b}\frac{\beta r^{2-\tilde{\alpha}}}{\left(N-1\right)\left(\tilde{\alpha}-2\right)}}}=e^{-\frac{\left(N-1\right)\left(\tilde{\alpha}-2\right)s}{2\pi\lambda_{b}\beta r^{2-\tilde{\alpha}}}}.
	\end{aligned}	
\end{equation}
}
Now we can write the probability in (\ref{eq:72}) as 
\begin{equation}
\textrm{Pr}\left(\gamma_{b}<\gamma_{r}\right)=\stackrel[i=0]{N-1}{\sum}\frac{\left(-s\right)^{i}}{i!}\frac{d^{i}}{ds^{i}}e^{-\frac{\left(N-1\right)\left(\tilde{\alpha}-2\right)s}{2\pi\lambda_{b}\beta r^{2-\tilde{\alpha}}}},
\end{equation}
which can be found as 
\vspace{-2mm}
\begin{equation}
\textrm{Pr}\left(\gamma_{b}<\gamma_{r}\right)=\stackrel[i=0]{N-1}{\sum}\frac{\left(-\frac{\alpha}{\gamma_{r}\left|\alpha_{b}\right|^{-2}}\right)^{i}}{i!}\left(-\frac{\left(N-1\right)\left(\tilde{\alpha}-2\right)}{2\pi\lambda_{b}\beta r^{2-\tilde{\alpha}}}\right)^{i}e^{-\frac{\left(N-1\right)\left(\tilde{\alpha}-2\right)\frac{\alpha}{\gamma_{r}\left|\alpha_{b}\right|^{-2}}}{2\pi\lambda_{b}\beta r^{2-\tilde{\alpha}}}},
\vspace{-2mm}
\end{equation}
and 
\vspace{-2mm}
\begin{equation}
\textrm{Pr}\left(\gamma_{b}>\gamma_{r}\right)=1-\stackrel[i=0]{N-1}{\sum}\frac{1}{i!}\left(-\frac{\alpha\left(N-1\right)\left(\tilde{\alpha}-2\right)\left|\alpha_{b}\right|^{2}}{2\pi\lambda_{b}\beta r^{2-\tilde{\alpha}}\gamma_{r}}\right)^{i}e^{-\frac{\left(N-1\right)\left(\tilde{\alpha}-2\right)\alpha\left|\alpha_{b}\right|^{2}}{2\pi\lambda_{b}\beta r^{2-\tilde{\alpha}}\gamma_{r}}}\label{eq:78}.
\vspace{-2mm}
\end{equation}

The parameter $\alpha_{b}$ is the reflection coefficient, which represents
the effects of both the radar path-loss and cross-section of the target,
thus it can be written as $\left|\alpha_{b}\right|^{2}=\tilde{\alpha_{b}}d_{t}^{-m}$
where $\tilde{\alpha_{b}}$ represent the effect of cross-section
of the target and $d_{t}^{-m}$ is the path-loss. {\textcolor{blue}{Since the distance
		between a point in $\mathbf{R}^{2}$ and the nearest BS is distributed
		as $f\left(x\right)=2\pi\lambda xe^{-2\pi\lambda x^{2}}$.}} The distance
between the served BS and the typical target can be written as $d_{t}=\sqrt{r_{t}^{2}+h_{t}^{2}}$
where $r_{t}$ is the horizontal distance and $h_{t}$ is the vertical
distance. The distribution of the distance between the served BS
and the typical target, $d_{t}$, for a given altitude can be derived
as 
\vspace{-2mm}
\begin{equation}
\textrm{F}_{d_{t}}\left(r_{r}\right)=\textrm{Pr}\left(\sqrt{r_{t}^{2}+h_{t}^{2}}<r\right)
\end{equation}
\begin{equation}
	\begin{aligned}
		\textrm{Pr}\left(\sqrt{r_{t}^{2}+h_{t}^{2}}<r\right)=\textrm{Pr}\left(r_{t}^{2}+h_{t}^{2}<r^{2}\right)=\textrm{Pr}\left(r_{t}<\sqrt{r^{2}-h_{t}^{2}}\right) =1-e^{-2\pi\lambda_{b}\left(r^{2}-h_{t}^{2}\right)}.
		\vspace{-2mm}
	\end{aligned}
\end{equation}
Thus we can find
\vspace{-2mm}
\begin{equation}
\textrm{F}_{d_{t}}\left(r\right)=1-e^{-2\pi\lambda_{b}\left(r^{2}-h_{t}^{2}\right)}\textrm{ and }f_{d_{t}}\left(r\right)=4r\pi\lambda_{b}e^{-2\pi\lambda_{b}\left(r^{2}-h_{t}^{2}\right)}.
\vspace{-2mm}
\end{equation}

Now, the coverage probability can be evaluated by taking the average
over the distance, $d_{t}$, thus (\ref{eq:78}) can be expressed
as
\[
\textrm{Pr}\left(\gamma_{b}>\gamma_{r}\right)=1-\stackrel[i=0]{N-1}{\sum}\frac{4\pi\lambda_{b}}{i!}\stackrel[0]{\infty}{\int}r\left(-\frac{\tilde{\alpha}\left(N-1\right)\left(\alpha-2\right)P_{b}\tilde{\alpha_{b}}r^{-\alpha_{r}}}{2\pi\lambda_{b}\beta r^{2-\alpha}\gamma_{r}}\right)^{i}
\]
\vspace{-2mm}
\begin{equation}
\times e^{-\left(\frac{\left(N-1\right)\left(\alpha-2\right)\alpha P_{b}\tilde{\alpha_{b}}r^{-\alpha_{r}}}{2\pi\lambda_{b}\beta r^{2-\alpha}\gamma_{r}}+2\pi\lambda_{b}\left(r^{2}-h_{t}^{2}\right)\right)}dr_{r}.
\vspace{-2mm}
\end{equation}
This expression can be simplified using Gaussian Quadrature rules
to 

\begin{align}
	\textrm{Pr}\left(\gamma_{b}>\gamma_{r}\right)=&1-\stackrel[i=0]{N-1}{\sum}\frac{4\pi\lambda_{b}}{i!}\stackrel[n=1]{Q}{\sum}\textrm{H}_{n}e^{r_{n}}r_{n}\left(-\frac{\tilde{\alpha}\left(N-1\right)\left(\alpha-2\right)P_{b}\tilde{\alpha_{b}}r_{n}^{-\alpha_{r}}}{2\pi\lambda_{b}\beta r^{2-\alpha}\gamma_{r}}\right)^{i} \nonumber \\
	&\times e^{-\left(\frac{\left(N-1\right)\left(\alpha-2\right)\alpha P_{b}\tilde{\alpha_{b}}r_{n}^{-\alpha_{r}}}{2\pi\lambda_{b}\beta r^{2-\alpha}\gamma_{r}}+2\pi\lambda_{b}\left(r_{n}^{2}-h_{t}^{2}\right)\right)}+R_{n}\label{eq:85},
\end{align}

\noindent where $r_{n}$ and $\textrm{H}_{n}$ are the $n^{th}$ zero
and the weighting factor of the Laguerre polynomials, respectively,
and the remainder $R_{n}$ is negligible for $Q>15$ \cite{book2}.
Finally, by substituting (\ref{eq:58}) into (\ref{eq:10}) we can
find the PSE of radar system presented in Theorem 2.

\bibliographystyle{IEEEtran}
\bibliography{bib}

\end{document}